\documentclass[
 reprint,
showpacs,
nofootinbib,
 amsmath,amssymb,
 aps
prd,
]{revtex4-1}

\usepackage{graphicx}
\usepackage{dcolumn}
\usepackage{bm}
\usepackage{aas_macros}
\usepackage{xfrac}
\usepackage{color}
\usepackage{subfigure}
\usepackage{enumitem} 
\usepackage{mathrsfs}
\usepackage{gensymb}
\usepackage{hyperref}

\newcommand\Tstrut{\rule{0pt}{3ex}}         
\newcommand\Bstrut{\rule[-2ex]{0pt}{0pt}}   

\begin{document}

\title{Integrated approach to cosmology: \\
Combining CMB, large-scale structure and weak lensing}

\author{Andrina Nicola}
\email{andrina.nicola@phys.ethz.ch} 
\affiliation{Department of Physics, ETH Zurich, Wolfgang-Pauli-Strasse 27, CH-8093 Zurich, Switzerland}
\author{Alexandre Refregier}
\affiliation{Department of Physics, ETH Zurich, Wolfgang-Pauli-Strasse 27, CH-8093 Zurich, Switzerland}
\author{Adam Amara}
\affiliation{Department of Physics, ETH Zurich, Wolfgang-Pauli-Strasse 27, CH-8093 Zurich, Switzerland}

\begin{abstract}

Recent observational progress has led to the establishment of the standard $\Lambda$CDM model for cosmology. This development is based on different cosmological probes that are usually combined through their likelihoods at the latest stage in the analysis. We implement here an integrated scheme for cosmological probes, which are combined in a common framework starting at the map level. This treatment is necessary as the probes are generally derived from overlapping maps and are thus not independent. It also allows for a thorough test of the cosmological model and of systematics through the consistency of different physical tracers. As a first application, we combine current measurements of the Cosmic Microwave Background (CMB) from the Planck satellite, and galaxy clustering and weak lensing from SDSS. We consider the spherical harmonic power spectra of these probes including all six auto- and cross-correlations along with the associated full Gaussian covariance matrix. This provides an integrated treatment of different analyses usually performed 
separately including CMB anisotropies, cosmic shear, galaxy clustering, galaxy-galaxy lensing and the Integrated Sachs-Wolfe (ISW) effect with galaxy and shear tracers. We derive constraints on $\Lambda$CDM parameters that are compatible with existing constraints and highlight tensions between data sets, which become apparent in this integrated treatment. We discuss how this approach provides
a complete and powerful integrated framework for probe combination and how it can be extended to include other tracers in the context of current and future wide field cosmological surveys.

\end{abstract}

\pacs{98.80.-k, 98.80.Es}
\maketitle

\section{\label{sec:intro} Introduction}

The past two decades have seen immense progress in observational cosmology that has lead to the establishment of the $\Lambda$CDM model for cosmology.
This development is mainly based on the combination of different cosmological probes such as the CMB temperature anisotropies, galaxy clustering, weak gravitational lensing, supernovae and galaxy clusters. Until now, these probes have been, for the most part, measured and analysed separately using different techniques and combined at late stages of the analysis, i.e. when deriving constraints on cosmological parameters. However, this approach is not ideal for current and future surveys such as the Dark Energy Survey (DES\footnote{\tt{http://www.darkenergysurvey.org}.}), the Dark Energy Spectroscopic Instrument (DESI\footnote{\tt{http://desi.lbl.gov}.}), the Large Synoptic Survey Telescope (LSST\footnote{\tt{http://www.lsst.org}.}), Euclid\footnote{\tt{http://sci.esa.int/euclid/}.} and the Wide Field Infrared Survey Telescope (WFIRST\footnote{\tt{http://wfirst.gsfc.nasa.gov}.}) for several reasons. First, these surveys will cover large, overlapping regions of the observable universe and are therefore not statistically independent. In addition, the analysis of these surveys requires tight control of systematic effects, which might be identified by a direct cross-correlation of the probes statistics. Moreover, each probe provides a measurement of the cosmic structures through a different physical field, such as density, velocity, gravitational potentials, and temperature. A promising way to test for new physics, such as modified gravity, is to look directly for deviations from the expected relationships of the statistics of the different fields. The integrated treatment of the probes from the early stages of the analysis will thus provide the cross-checks and the redundancy needed not only to achieve high-precision but also to challenge the different sectors of the cosmological model.

Several earlier studies have considered joint analyses of various cosmological probes. \citet{Mandelbaum:2013, Cacciato:2013} and \citet{Kwan:2016} for example derived cosmological constraints from a joint analysis of galaxy-galaxy lensing and galaxy clustering while \citet{Liu:2016} used the cross-correlation between the galaxy shear field and the overdensity field together with the cross-correlation of the galaxy overdensity with CMB lensing to constrain multiplicative bias in the weak lensing shear measurement in CFHTLenS. Recently, \citet{Singh:2016} performed a joint analysis of CMB lensing as well as galaxy clustering and weak lensing. Furthermore, \citet{Eifler:2014} and \citet{Krause:2016} have theoretically investigated joint analyses for photometric galaxy surveys by modelling the full non-Gaussian covariance matrix between cosmic shear, galaxy-galaxy lensing, galaxy clustering, photometric baryon acoustic oscillations (BAO), galaxy cluster number counts and galaxy cluster weak lensing.

Extending beyond this, we present and implement an integrated approach to probe combination. In this first implementation we combine data from CMB temperature anisotropies, galaxy overdensities and weak lensing. We use data from Planck 2015 \cite{Planck-Collaboration:2015af} for the CMB, for galaxy clustering we use photometric data from the 8$^{\mathrm{th}}$ data release of the Sloan Digital Sky Survey (SDSS DR 8) \cite{Aihara:2011} and the weak lensing shear data comes from SDSS Stripe 82 \cite{Annis:2014}. We combine these probes into a common framework at the map level by creating projected 2-dimensional maps of CMB temperature, galaxy overdensity and the weak lensing shear field. In order to jointly analyse this set of maps we consider the spherical harmonic power spectra of the probes including their cross-correlations. This leads to a spherical harmonic power spectrum matrix that combines CMB temperature anisotropies, galaxy clustering, cosmic shear, galaxy-galaxy lensing and the ISW \cite{Sachs:1967} effect with galaxy and weak lensing shear tracers. We combine this power spectrum matrix together with the full Gaussian covariance matrix and derive constraints on the parameters of the $\Lambda$CDM cosmological model, marginalising over a constant linear galaxy bias and a parameter accounting for possible multiplicative bias in the weak lensing shear measurement. In this first implementation, we use some conservative and simplifying assumptions. For instance we include a limited range of angular scales for the different probes to reduce our sensitivity to systematics, nuisance parameters and nonlinear corrections. With this, we work under the assumption of Gaussian covariance matrices and with a reduced set of nuisance parameters. 
 
This paper is organised as follows. In Section \ref{sec:framework} we describe the framework for integrated probe combination employed in this work. The theoretical modelling of the cosmological observables is summarised in Section \ref{sec:theorypred}. Section \ref{sec:maps} describes the data analysis for each probe, especially the map-making procedure. The computation of the spherical harmonic auto- and cross-power spectra is discussed in Section \ref{sec:cls} and the estimation of the covariance matrix is detailed in Section \ref{sec:covariance}. In Section \ref{sec:results} we present the cosmological constraints derived from the joint analysis and we conclude in Section \ref{sec:conclusions}. More detailed descriptions of data analysis as well as robustness tests are deferred to the Appendix.

\section{\label{sec:framework} Framework}

\begin{figure}
\begin{center}
\includegraphics[scale=0.55]{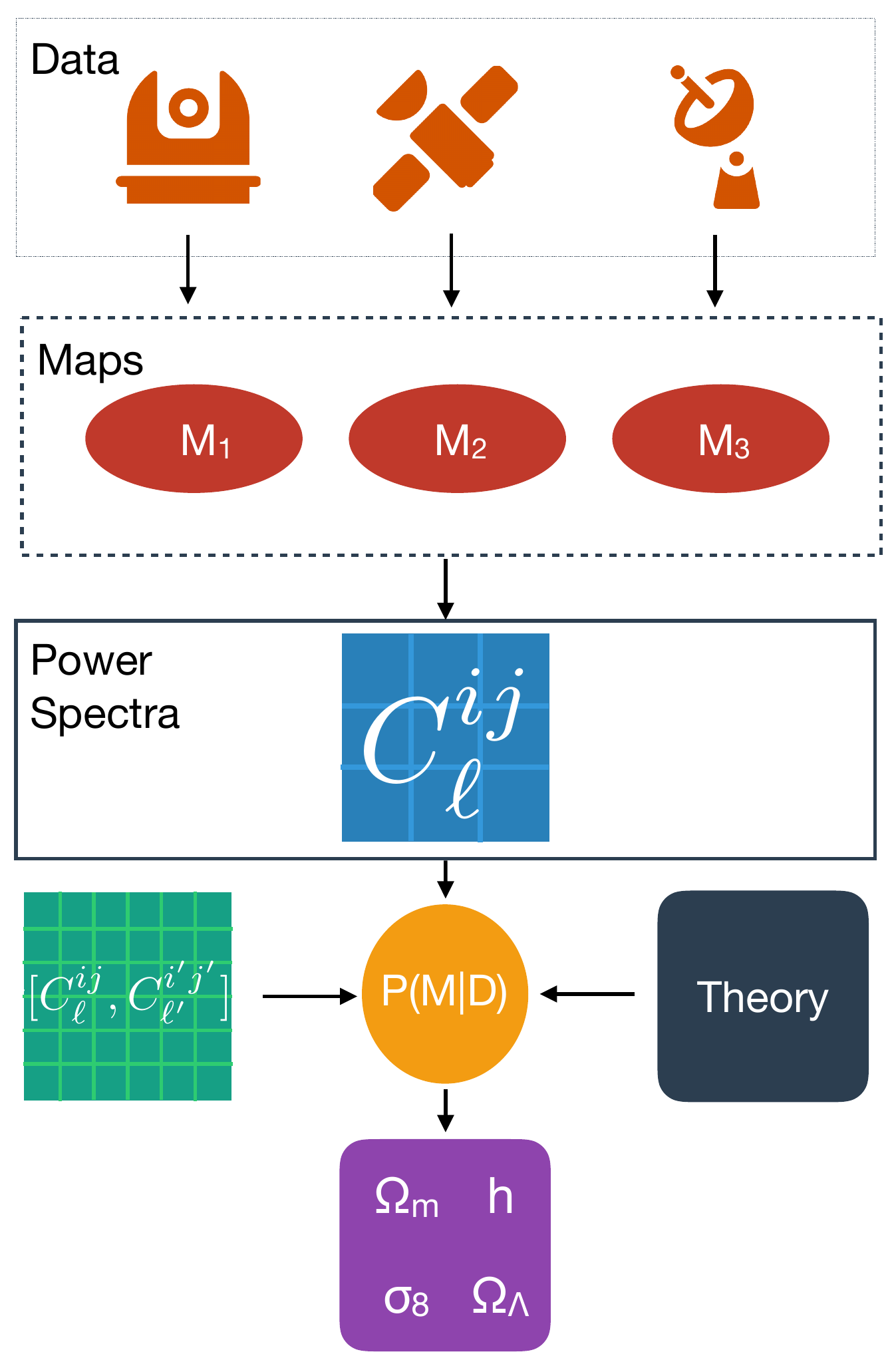}
\caption{Synopsis of the framework for integrated probe combination employed in this work.}
\label{fig:framework}
\end{center}
\end{figure}

The framework for integrated probe combination employed in this work is illustrated in Fig.~\ref{fig:framework}. In a first step we collect data for different cosmological probes as taken by either separate surveys or by the same survey. For our first implementation described below we use cosmological data from the CMB temperature anisotropies, the galaxy overdensity field and the weak lensing shear field. After data collection, we perform probe specific data analysis which involves data selection and systematics removal. We then homogenise the data format by creating projected 2-dimensional maps for all probes considered. The common data format allows us to combine the cosmological probes into a common framework at the map level. We compute both the spherical harmonic auto- and cross-power spectra of this set of maps and combine them into the spherical harmonic power spectrum matrix $C_{\ell}^{ij}$. This matrix captures the cosmological information contained in the two-point statistics of the maps. In a last step we compute the power spectrum covariance matrix and combine it with theoretical predictions to derive constraints on cosmological parameters from a joint fit to the measured spherical harmonic power spectra. The details of the implementation for CMB temperature anisotropies, galaxy overdensities and weak lensing are described below. 

\section{\label{sec:theorypred} Theoretical predictions}

The statistical properties of both galaxy overdensity $\delta_{g}$ and weak lensing shear $\gamma$, as well as their cross-correlation can be measured from their spherical harmonic power spectra. These generally take the form of weighted integrals of the nonlinear matter power spectrum $P^{\mathrm{nl}}_{\delta \delta}(k, z)$ multiplied with spherical Bessel functions $j_{\ell}\boldsymbol{(}k \chi(z)\boldsymbol{)}$. Their computation is time-consuming and we therefore resort to the the Limber approximation \cite{Limber:1953, Kaiser:1992, Kaiser:1998} to speed up calculations. This is a valid approximation for small angular scales, typically $\ell > \mathcal{O}(10)$, and broad redshift bins \cite{Peacock:1999}. For simplicity, we further focus on flat cosmological models, i.e. $\Omega_{\mathrm{k}} = 0$, for the theoretical predictions. The spherical harmonic power spectrum $C_{\ell}^{ij}$ at multipole $\ell$ between cosmological probes $i$, $j$ $\in \{\delta_{g}, \gamma\}$ can then be expressed as:
\begin{multline}
C_{\ell}^{ij}=\int \mathrm{d} z \; \frac{c}{H(z)} \; \frac{W^{i}\boldsymbol{\left(}\chi(z)\boldsymbol{\right)}W^{j}\boldsymbol{\left(}\chi(z)\boldsymbol{\right)}}{\chi^{2}(z)} \\
\times P^{\mathrm{nl}}_{\delta \delta}\left(k=\frac{\ell+\sfrac{1}{2}}{\chi(z)}, z\right),
\end{multline}
where $c$ is the speed of light, $\chi(z)$ the comoving distance, $H(z)$ the Hubble parameter and $W^{i'}\boldsymbol{\left(}\chi(z)\boldsymbol{\right)}$ denotes the window function for probe $i'$.

For galaxy clustering the window function is given by
\begin{equation}
W^{\delta_{g}}\boldsymbol{\left(}\chi(z)\boldsymbol{\right)} = \frac{H(z)}{c} b(z) n(z),
\label{eq:deltagwindow}
\end{equation}
where $b(z)$ denotes a linear galaxy bias and $n(z)$ is the normalised redshift selection function of the survey i.e. $\int \mathrm{d}z \; n(z) = 1$. We focus on scale-independent galaxy bias since we restrict the analysis to large scales, which are well-described by linear theory. 

The window function for weak lensing shear is
\begin{equation}
W^{\gamma}\boldsymbol{\left(}\chi(z)\boldsymbol{\right)} = \frac{3}{2} \frac{\Omega_{\mathrm{m}} H^{2}_{0}}{c^{2}} \frac{\chi(z)}{a} \int_{\chi(z)}^{\chi_{\mathrm{h}}} \mathrm{d} z' n(z') \frac{\chi(z')-\chi(z)}{\chi(z')},
\label{eq:gammawindow}
\end{equation}
where $\Omega_{\mathrm{m}}$ denotes the matter density parameter today, $H_{0}$ is the present-day Hubble parameter, $\chi_{\mathrm{h}}$ is the comoving distance to the horizon and $a$ denotes the scale factor.

Similarly to the spherical harmonic power spectra of galaxy clustering and weak lensing the spherical harmonic power spectrum of CMB temperature anisotropies $T$ can be related to the primordial matter power spectrum generated during inflation as \cite{Dodelson:2003}
\begin{equation}
C^{\mathrm{TT}}_{\ell} = \frac{2}{\pi} \int \mathrm{d}k \; k^{2} P^{\mathrm{lin}}_{\delta \delta}(k) \left \vert \frac{\Delta T_{\ell}(k)}{\delta(k)}\right \vert^{2},
\end{equation}
where $\Delta T_{\ell}$ denotes the transfer function of the temperature anisotropies and $\delta$ is the matter overdensity.

The CMB temperature anisotropies are correlated to tracers of the large-scale structure (LSS) such as galaxy overdensity and weak lensing shear primarily through the integrated Sachs-Wolfe effect \cite{Sachs:1967}. On large enough scales where linear theory holds, the spherical harmonic power spectra between these probes can be computed from expressions similar to those above. In the Limber approximation \cite{Limber:1953, Kaiser:1992, Kaiser:1998}, the spherical harmonic power spectrum between CMB temperature anisotropies and a tracer $i$ of the LSS becomes: \cite{Crittenden:1996}
\begin{multline}
C^{i\mathrm{T}}_{\ell} = 3 \frac{\Omega_{\mathrm{m}} H^{2}_{0}T_{\mathrm{CMB}}}{c^{2}} \frac{1}{(\ell+\sfrac{1}{2})^{2}} \int \mathrm{d}z \frac{\mathrm{d}}{\mathrm{d}z} \left[D(z)(1+z)\right] \\
\times D(z) W^{i}\boldsymbol{\left(}\chi(z)\boldsymbol{\right)} P^{\mathrm{lin}}_{\delta \delta}\left(k=\frac{\ell+\sfrac{1}{2}}{\chi(z)}, 0\right),
\label{eq:clisw}
\end{multline}
where $T_{\mathrm{CMB}}$ denotes the mean temperature of the CMB today, $i \in \{\delta_{g}, \gamma\}$ and $W^{i}\boldsymbol{\left(}\chi(z)\boldsymbol{\right)}$ represents the window functions defined in Equations \ref{eq:deltagwindow} and \ref{eq:gammawindow}. We have further split the linear matter power spectrum $P^{\mathrm{lin}}_{\delta \delta}(k, z)$ into its time-dependent part parametrised by the growth factor $D(z)$ and the scale-dependent part $P^{\mathrm{lin}}_{\delta \delta}(k, 0)$. For a derivation of Eq.~\ref{eq:clisw} for the galaxy overdensity field as tracer of the LSS see e.g. \citet{Padmanabhan:2005}. The derivation for $C^{\gamma\mathrm{T}}_{\ell}$ is similar and is detailed in Appendix \ref{sec:shearisw}.

To compute the auto-power spectrum of the CMB temperature anisotropies we use the publicly available Boltzmann code $\textsc{class}$\footnote{$\tt{http://class\text{-}code.net}$.} \cite{Lesgourgues:2011}. For the other power spectra we use $\textsc{PyCosmo}$ \cite{Refregier:2016}. We calculate the linear matter power spectrum from the transfer function derived by \citet{Eisenstein:1998}. To compute the nonlinear matter power spectrum we use the $\textsc{Halofit}$ fitting function \cite{Smith:2003} with the revisions of \citet{Takahashi:2012}. 

\section{\label{sec:maps} Maps}

\begin{table*}
\caption{Summary of used data.} \label{tab:data}
\begin{center}
\begin{tabular}{>{\centering}m{2.5cm}|>{\centering}m{7cm}|>{\centering}m{8cm}@{}m{0pt}@{}} \hline \hline
                                 
CMB temperature anisotropies & 
\multicolumn{2}{>{\centering}m{8cm}}{
\Tstrut 
Survey: Planck 2015 \cite{Planck-Collaboration:2015ab} \\
Fiducial foreground-reduced map: $\tt{Commander}$ \\
Sky coverage: $f_{\text{sky}} = 0.776$
\Bstrut} &

\tabularnewline \hline
galaxy overdensity & 
\multicolumn{2}{>{\centering}m{8cm}}{
\Tstrut 
Survey: SDSS DR8 \cite{Aihara:2011} \\
Sky coverage: $f_{\mathrm{sky}} = 0.27$ \\
Galaxy sample: CMASS1-4 \\
Number of galaxies: $N_{\mathrm{gal}} = 854\,063$ \\
Photometric redshift range $0.45 \leq z_{\mathrm{phot}} < 0.65$ 
\Bstrut} &

\tabularnewline \hline
weak lensing & 
\multicolumn{2}{>{\centering}m{8cm}}{
\Tstrut 
Survey: SDSS Stripe 82 co-add \cite{Annis:2014}\\
Sky coverage: $f_{\mathrm{sky}} = 0.0069$ \\
Number of galaxies: $N_{\mathrm{gal}} = 3\,322\,915$ \\
Photometric redshift range: $0.1 \lesssim z_{\mathrm{phot}} \lesssim 1.1$ \\
r.m.s. ellipticity per component: $\sigma_{e} \sim 0.43$
\Bstrut} &

\tabularnewline \hline \hline

\end{tabular}
\end{center}
\end{table*} 

\subsection{\label{subsec:cmbmap}Cosmic Microwave Background}

\begin{figure*}
\begin{center}
\includegraphics[scale=0.7]{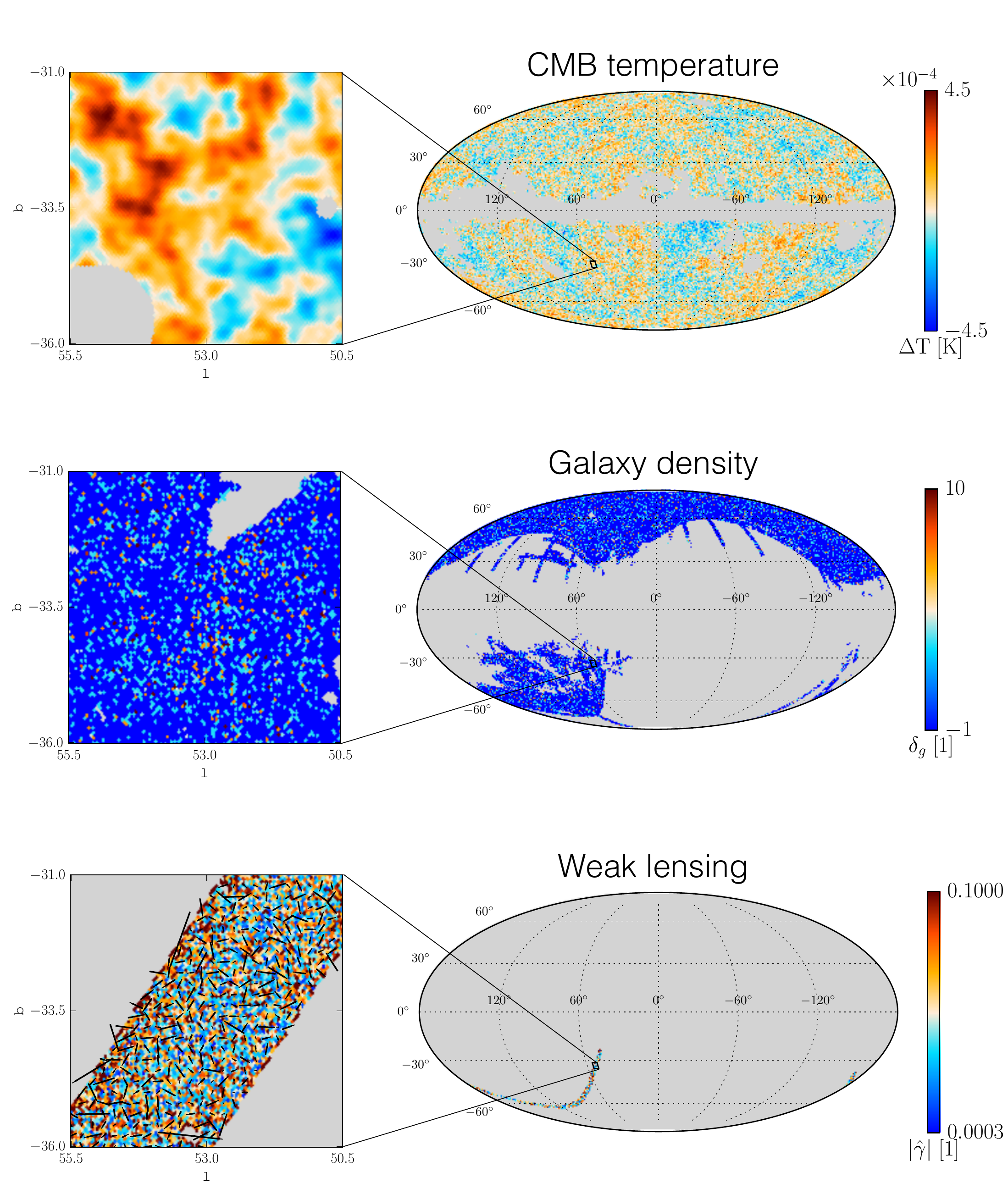}
\caption{Summary of the three maps in Galactic coordinates used in this analysis. The all-sky maps are in Mollweide projection while the zoom-in versions are in Gnomonic projection. The HMHS map of CMB temperature anisotropies as derived using $\tt{Commander}$ is shown in the top panel. It is masked using the $\tt{UT78}$ mask. The middle panel shows the systematics-corrected (see text) galaxy overdensity map for CMASS1-4 galaxies. Grey areas have been masked either because they lie outside the survey footprint or are potentially contaminated by systematics. The lower panel shows the map of the SDSS Stripe 82 shear modulus $\vert \hat{\gamma} \vert$. Grey areas have been masked because they are either unobserved or do not contain galaxies for shear measurement. The zoom-in figures (left) are enlarged versions of the $5\times5$ deg$^{2}$ region centred on $(\tt{l}, \tt{b})$ $= (53\degree, -33.5\degree)$ shown in the maps. The zoom-in for the galaxy shear map is overlaid with a whisker plot of the galaxy shears. All three maps have resolution $\tt NSIDE$ $= 1024$.}
\label{fig:maps}
\end{center}
\end{figure*}

We use the foreground-reduced CMB anisotropy maps provided by the Planck collaboration \cite{Planck-Collaboration:2015ab} in their 2015 data release. We choose these over the uncleaned single-frequency maps because they allow to perform the foreground correction on the maps rather than the power spectrum level. This is important when considering probe combination. The Planck foreground-reduced CMB anisotropy maps have been derived using four different algorithms: $\tt Commander$, $\tt NILC$, $\tt SEVEM$ and $\tt SMICA$. The maps are given in HEALPix\footnote{$\tt http://healpix.sourceforge.net$.} \cite{Gorski:2005} format and are provided in Galactic coordinates at two different resolutions of $\tt NSIDE$ $= 1024$ and $\tt NSIDE$ $= 2048$. These correspond to pixel areas of $11.8$ and $2.95$ arcmin$^{2}$ respectively. Different data configurations are available \cite{Planck-Collaboration:2015ab}; we use both the half-mission half-sum (HMHS) maps, which contain both signal and noise, and the half-mission half-difference maps (HMHD), which contain only noise and potential residual systematic uncertainties. All four maps yield consistent estimates of both the spherical harmonic power spectrum of the CMB temperature anisotropies as well as the spherical harmonic cross-power spectrum between CMB temperature anisotropies and tracers of the LSS \cite{Planck-Collaboration:2015ab, Planck-Collaboration:2015ac, Planck-Collaboration:2015ad}. Since the Planck collaboration found the $\tt Commander$ approach to be the preferred solution for studying the CMB anisotropies at large and intermediate angular scales, we also choose it for our analysis. Each of the four foreground reduction methods also provides a confidence mask inside which the CMB solution is trusted. Following the Planck collaboration \cite{Planck-Collaboration:2015ab}, we adopt the union of the confidence masks for $\tt Commander$, $\tt SEVEM$ and $\tt SMICA$. This is referred to as the $\tt UT78$ mask and covers $77.6 \%$ of the sky at a resolution of $\tt NSIDE$ $= 2048$. To downgrade the mask to $\tt NSIDE$ $= 1024$, we follow the description outlined in \citet{Planck-Collaboration:2015ab}. The HMHS CMB anisotropy map derived using $\tt Commander$ is shown in the top panel of Fig.~\ref{fig:maps} for resolution $\tt NSIDE$ $= 1024$ and the corresponding HMHD map is shown in Fig.~\ref{fig:cmbnoisemap} in the Appendix.

\subsection{\label{subsec:galaxymap}Galaxy overdensity}

The SDSS \cite{York:2000, Eisenstein:2011, Gunn:1998, Gunn:2006} obtained wide-field images of $14\,555$ deg$^{2}$ of the sky in 5 photometric passbands ($u, g, r, i, z$ \cite{Fukugita:1996, Smith:2002, Doi:2010}) up to a limiting $\tt{r\text{-}band}$ magnitude of $r \simeq 22.5$. The photometric data is complemented with spectroscopic data from the Baryonic Oscillations Spectroscopic Survey (BOSS) \cite{Eisenstein:2011, Dawson:2013, Smee:2013}. BOSS was conducted as part of SDSS III \cite{Eisenstein:2011} and obtained spectra of approximately 1.5 million luminous galaxies distributed over $10\,000$ deg$^{2}$ of the sky. The SDSS photometric redshifts for DR8 \cite{Aihara:2011} are estimated using a local regression model trained on a spectroscopic training set consisting of $850\,000$ SDSS DR8 spectra and spectroscopic data from other surveys\footnote{More details can be found on \tt{http://www.sdss3.org/dr8/algorithms/photo-z.php}.}. The algorithm is outlined in \citet{Beck:2016}. 

In our analysis, which is described in the following, we largely follow \citet{Ho:2012}. We select objects classified as galaxies from the $\textsc{PhotoPrimary}$ table in the Catalog Archive Server (CAS\footnote{The SDSS Catalog Archive Server can be accessed through \tt{http://skyserver.sdss.org/CasJobs/SubmitJob.aspx}.\label{footn:cas}}). To obtain a homogeneous galaxy sample we further select CMASS galaxies using the color-magnitude cuts used for BOSS target selection \cite{Eisenstein:2011} and outlined in \citet{Ho:2012}. This selection isolates luminous, high-redshift galaxies that are approximately stellar mass limited \cite{White:2011, Ross:2011}. We further restrict the sample to CMASS galaxies with SDSS photometric redshifts between $0.45 \leq z < 0.65$, i.e. we consider the photometric redshift slices CMASS1-4. This selection yields a total of $N_{\mathrm{gal}} = 1\,096\,455$ galaxies.

To compute the galaxy overdensity field, we need to characterise the full area observed by the survey and mask regions heavily affected by foregrounds or potential systematics. The area imaged by the SDSS is divided into units called fields. Several such fields have been observed multiple times in the SDSS imaging runs. The survey footprint is the union of the best observed (primary) fields at each position and is described in terms of $\textsc{Mangle}$ \cite{Hamilton:1993, Hamilton:2004, Swanson:2008} spherical polygons. Each of these polygons is matched to the SDSS field fully covering it\footnote{This information is found in the files $\tt window\_unified.fits$ and $\tt window\_flist.fits$.}. In order to select the survey area least affected by foregrounds and potential systematics we follow \citet{Ho:2012} and \citet{Ross:2011} and restrict the analysis to polygons covered by fields with $\tt{score}$\footnote{\tt{http://www.sdss3.org/dr10/algorithms/resolve.php}.} $\geq 0.6$, full width at half maximum (FWHM) of the point spread function (PSF) $\tt{PSF\text{-}FWHM}$ $< 2.0$ arcsec in the $\tt{r\text{-}band}$ and Galactic extinction $\tt{E(B-V)}$ $\leq 0.08$ as determined from the extinction maps from \citet{Schlegel:1998}. 

To facilitate a joint analysis between the LSS probes and the CMB, which is given as a map in Galactic coordinates, we transform both the galaxy positions as well as the survey mask from equatorial ($\tt RA$, $\tt DEC$) to Galactic ($\tt l$, $\tt b$) coordinates. We construct the continuous galaxy overdensity field by pixelising the galaxy overdensities $\delta_{g} = \delta n/\bar{n}$ onto a HEALPix pixelisation of the sphere with resolution $\tt NSIDE$ $= 1024$. We mask the galaxy overdensity map with a HEALPix version of the SDSS survey mask, which is obtained by random sampling of the $\textsc{Mangle}$ mask. To account for the effect of bright stars, we use the Tycho astrometric catalog \cite{Hog:2000} and define magnitude-dependent stellar masks as defined in \citet{Padmanabhan:2007}. We remove galaxies inside the bright star masks and correct for the area covered by the bright stars by removing the area covered by the star masks from the pixel area $A_{\mathrm{pix}, \mathrm{corr}} = A_{\mathrm{pix}, \mathrm{uncorr}} - A_{\mathrm{stars}}$ when computing the galaxy overdensity. The final map covers a fraction $f_{\mathrm{sky}} \approx 0.27$ of the sky and contains $N_{\mathrm{gal}} = 854\,063$ galaxies. 

Even after masking and removal of high contamination regions, there are still systematics left in the galaxy overdensity map. The correction for residual systematic uncertainties in the maps follows \citet{Ross:2011} and \citet{Hernandez-Monteagudo:2014} and is described in Appendix \ref{sec:galaxysys}. The final map is shown in the middle panel of Fig.~\ref{fig:maps}.

\begin{figure}
\begin{center}
\includegraphics[scale=0.6]{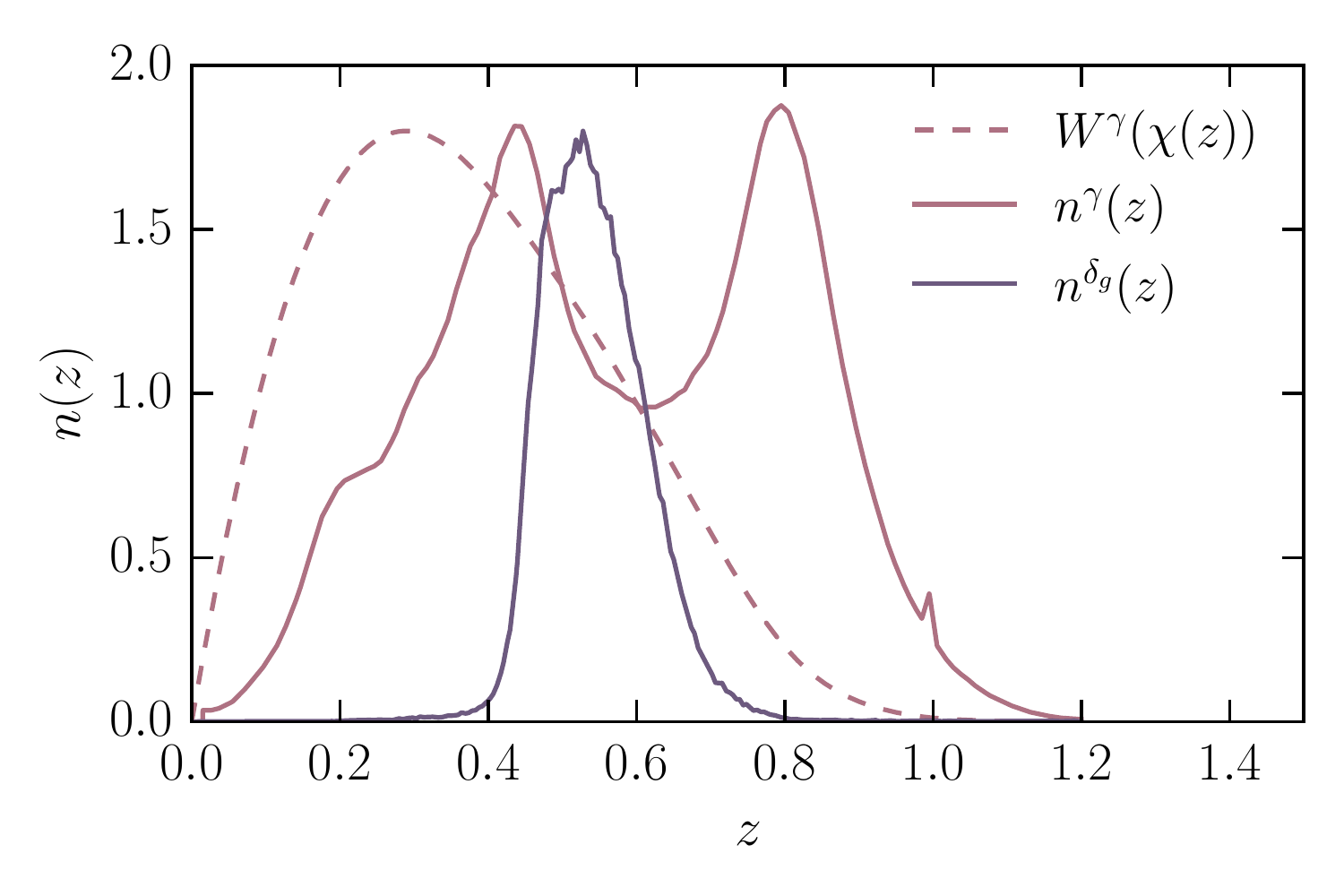}
\caption{Redshift distribution for the LSS probes. The figure shows the redshift selection function of SDSS CMASS1-4 galaxies, the redshift selection function for the SDSS Stripe 82 galaxies as well as the weak lensing shear window function defined in Eq.~\ref{eq:gammawindow}. The redshift selection function for CMASS1-4 galaxies as well as the weak lensing shear window function have been rescaled relative to the Stripe 82 redshift selection function.}
\label{fig:nz}
\end{center}
\end{figure}

As well as the maps we need an estimate for the redshift distribution of the galaxies in our sample. To this end we follow \citet{Ho:2012} and match photometrically detected galaxies to galaxies observed spectroscopically in SDSS DR9 \cite{Ahn:2012}. We then estimate the redshift distribution of the photometric galaxies from the spectroscopic redshift distribution of the matching galaxies. The selected CMASS1-4 galaxies have spectroscopic redshifts $0.4 \lesssim z \lesssim 0.7$ as can be seen from the redshift distribution shown in Fig.~\ref{fig:nz}.

\subsection{\label{subsec:ellipticitymap}Weak lensing}

We take weak lensing data from the SDSS Stripe 82 co-add \cite{Annis:2014}, which comprises $275$ deg$^{2}$ of co-added SDSS imaging data with a limiting $\tt{r\text{-}band}$ magnitude $r \approx 23.5$ and $\tt{r\text{-}band}$ median seeing of $1.1$ arcsec. The shapes of objects detected in the SDSS were measured from the adaptive moments \cite{Bernstein:2002} by the $\tt PHOTO$ pipeline \cite{Lupton:2001} and are available on the CAS\footnote{See footnote \ref{footn:cas}.}. Photometric redshifts for all detected galaxies were computed using a neural network approach as described in \citet{Reis:2012} and are available as a DR7 value added catalog\footnote{$\tt{http://classic.sdss.org/dr7/products/value\_added/}$, $\tt{http://das.sdss.org/va/coadd\_galaxies/}$.}. 

In the following analysis we closely follow the work by \citet{Lin:2012}. We select objects identified as galaxies in the co-add data (i.e. $\tt run = 106$ or $\tt run = 206$) from the CAS and we restrict the sample to galaxies with extinction corrected $\tt{i\text{-}band}$ magnitudes in the range $18 < i < 24$. Further we select only objects that pass the clean photometry cuts as defined by the SDSS\footnote{\tt{http://www.sdss.org/dr12/tutorials/flags/}.\label{footn:flags}} and do not have flags indicating problems with the measurement of adaptive moments as well as negative errors on those. The former cuts especially exclude galaxies containing saturated pixels. We use shapes measured in the $\tt{i\text{-}band}$ since it has the smallest seeing ($1.05$ arcsec) \cite{Annis:2014, Lin:2012} and further consider only galaxies with observed sizes at least $50 \%$ larger than the PSF. This requirement is quantified by requiring the resolution factor $R = 1-\tt mRrCcPSF/\tt mRrCc$ \cite{Bernstein:2002} to satisfy $R > 0.33$, where $\tt mRrCc$ and $\tt mRrCcPSF$ denote the sum of the second order moments in the CCD column and row direction for both the object and the PSF.
 
For the above galaxy sample we compute PSF-corrected galaxy ellipticities using the linear PSF correction algorithm as described in \citet{Hirata:2003}. For weak lensing shear measurement we follow \citet{Lin:2012} and restrict the sample to galaxies with PSF-corrected ellipticity components $e_{1}, e_{2}$ satisfying $\vert e_{1} \vert < 1.4$ as well as $\vert e_{2} \vert < 1.4$ and photometric redshift uncertainties $\sigma_{z} < 0.15$. This sample has an r.m.s. ellipticity per component of $\sigma_{e} \sim 0.43$. We then turn the PSF-corrected ellipticities for this sample into shear estimates. The details of the analysis are described in Appendix \ref{sec:cosmicshearanalysis}.

After computing weak lensing shear estimates from the ellipticities we apply a rotation to both the galaxy positions and shears from equatorial to Galactic coordinates\footnote{The exact rotation of the shears is described in Appendix \ref{sec:shearrotation}.} to allow for combination with the CMB. We pixelise both weak lensing shear components onto separate HEALPix pixelisations of the sphere choosing a resolution of $\tt NSIDE$ $= 1024$ as for the galaxy overdensity map. At this resolution the mean number of galaxies per pixel is about $38$, which corresponds to $n_{\mathrm{gal}} \simeq 3.2$ arcmin$^{-2}$. We apply a mask to both maps, which accounts for both unobserved and empty pixels. The final maps are constructed using $N_{\mathrm{gal}} =  3\,322\,915$ galaxies and cover a sky fraction $f_{\mathrm{sky}} \approx 0.0069$. The map of the shear modulus $\vert \hat{\gamma} \vert$ is shown in the bottom panel of Fig.~\ref{fig:maps} together with a zoom-in region with overlaid whisker plot illustrating the magnitude and direction of the weak lensing shear.

We follow \citet{Lin:2012} and estimate the redshift distribution of the galaxies from their photometric redshift distribution. The redshift distribution is shown in Fig.~\ref{fig:nz} together with the window function for weak lensing shear defined in Eq.~\ref{eq:gammawindow}. We see that the selected galaxies have photometric redshifts $z \lesssim 1.0$.

\section{\label{sec:cls} Spherical harmonic power spectra}

We calculate the spherical harmonic power spectra of the maps presented in the previous section using the publicly available code $\tt PolSpice$\footnote{\tt{http://www2.iap.fr/users/hivon/software/PolSpice/}.} \cite{Szapudi:2001, Chon:2004}. The $\tt PolSpice$ code is designed to combine both real and Fourier space in order to correct spherical harmonic power spectra measured on a cut-sky from the effect of the mask. The algorithm can be summarised as follows: starting from a masked HEALPix map, $\tt PolSpice$ first computes the so-called pseudo power spectrum, which is then Fourier transformed to the real space correlation function. In order to correct for the effects of the mask, the latter is divided by the mask correlation function. In a last step, the demasked correlation function is Fourier transformed back to the spherical harmonic power spectrum. This approach ensures that $\tt PolSpice$ can exploit the advantages of real space while still performing the computationally expensive calculations in Fourier space.

Demasking can only be performed on angular scales on which information is available, which translates to a maximal angular scale $\theta_{\mathrm{max}}$ for which a demasked correlation function can be computed. This maximal scale leads to ringing when transforming back from real to Fourier space, which can be reduced by apodising the correlation function prior to inversion. Both these steps lead to biases in the power spectrum recovered by $\tt PolSpice$. The kernels relating the average $\tt PolSpice$ estimates to the true power spectra can be computed theoretically for a given maximal angular scale and apodisation prescription and need to be corrected for when comparing theoretical predictions to observed power spectra.

An additional difficulty arises in the computation of spherical harmonic power spectra of spin-2 fields. Finite sky coverage tends to cause mixing between E-and B-modes. The polarisation version of $\tt PolSpice$ is designed to remove E- to B-mode leakage in the mean \cite{Chon:2004}. Details on our earlier application of $\tt PolSpice$ to LSS data are described in Appendix A of \citet{Becker:2015}. 

In order to calculate both the auto- and cross-power spectra for all probes, we need to estimate the maximal angular scale $\theta_{\mathrm{max}}$. This is not a well-defined quantity but we can separately estimate it for each probe from the real space correlation function of its mask. The real space correlation function of the survey mask will fall off significantly or vanish for scales larger than $\theta_{\mathrm{max}}$. We therefore estimate $\theta_{\mathrm{max}}$ as the scale around which the mask correlation function significantly decreases in amplitude. Appendix \ref{sec:maskcorr} illustrates this analysis for the example of the SDSS Stripe 82 weak lensing shear mask. In order to reduce Fourier ringing we apodise the correlation function using a Gaussian window function; following \citet{Chon:2004} we choose the FWHM of the Gaussian window as $\theta_{\mathrm{FWHM}} = \sfrac{\theta_{\mathrm{max}}}{2}$. Survey masks with complicated angular dependence might not exhibit a clear fall-off, which complicates the choice of $\theta_{\mathrm{max}}$. We therefore validate our choices of $\theta_{\mathrm{max}}$ and $\theta_{\mathrm{FWHM}}$ with the Gaussian simulations as described in Appendix \ref{sec:corrmaps} and \ref{sec:validation}. We find our choices to allow the recovery of the input power spectra for all the probes and settings.

All spherical harmonic power spectra are corrected for the effect of the HEALPix pixel window function and the power spectra involving the CMB map are further corrected for the Planck effective beam window function, which complements the CMB maps.

We now separately describe the measurement of all the six spherical harmonic power spectra. To compute the power spectra, we use the maps and masks described in Section \ref{sec:maps} at resolution $\tt NSIDE$ $= 1024$, except for the CMB temperature power spectrum. For the latter we use the maps at resolution $\tt NSIDE$ $= 2048$, but we do not expect this to make a significant difference.  The $\tt{PolSpice}$ parameter settings used to compute the power spectra are summarised in Tab.~\ref{tab:clparams}. This table further gives the angular multipole range as well as binning scheme employed for the cosmological analysis. For all probes considered, the uncertainties are derived from the Gaussian simulations described in Section \ref{sec:mockcov} and Appendix \ref{sec:corrmaps}. 

\begin{table}
\caption{Spherical harmonic power spectrum parameters and angular multipole ranges.} \label{tab:clparams}
\begin{center}
\begin{ruledtabular}
\begin{tabular}{ccccc}
Power spectrum & $\theta_{\mathrm{max}}$ [deg] & $\theta_{\mathrm{FWHM}}$ [deg] & $\ell$-range & $\Delta \ell$ \\ \hline \Tstrut                             
$C^{\mathrm{TT}}_{\ell}$ & 40 & 20 & $[10, \,610]$ & 30  \\
$C^{\delta_{g} \delta_{g}}_{\ell}$ & 80 & 40 & $[30, \,210]$ & 30  \\
$C^{\gamma \gamma}_{\ell}$ & 10 & 5 & $[70, \,610]$ & 60  \\
$C^{\delta_{g}\mathrm{T}}_{\ell}$ & 40 & 20 & $[30, \,210]$ & 30  \\
$C^{\gamma \mathrm{T}}_{\ell}$ & 10 & 5 & $[70, \,610]$ & 60  \\
$C^{\gamma \delta_{g}}_{\ell}$ & 10 & 5 & $[30, \,210]$ & 60  \\
\end{tabular}
\end{ruledtabular}
\end{center}
\end{table} 

\subsection{\label{subsec:tempcls} CMB}

We use the half-mission half-sum (HMHS) map to estimate the CMB signal power spectrum and the half-mission half-difference (HMHD) map to estimate the noise in the power spectrum of the HMHS map. 

The minimal angular multipole used in the cosmological analysis is chosen such as to minimise demasking effects and the cut at $\ell = 610$ ensures that we are not biased by residual foregrounds in the maps as discussed in Section \ref{sec:results}. The resulting power spectrum is shown in the top panel of Fig.~\ref{fig:cls}. In Appendix \ref{sec:cltests} we compare the CMB auto-power spectrum computed from the different foreground-reduced maps. As illustrated in Fig.~\ref{fig:compplanckmaps} in Appendix \ref{sec:cltests} we find that the measured CMB auto-power spectrum is unaffected by the choice of foreground-reduced map. 

\begin{figure*}
\includegraphics[scale=0.4]{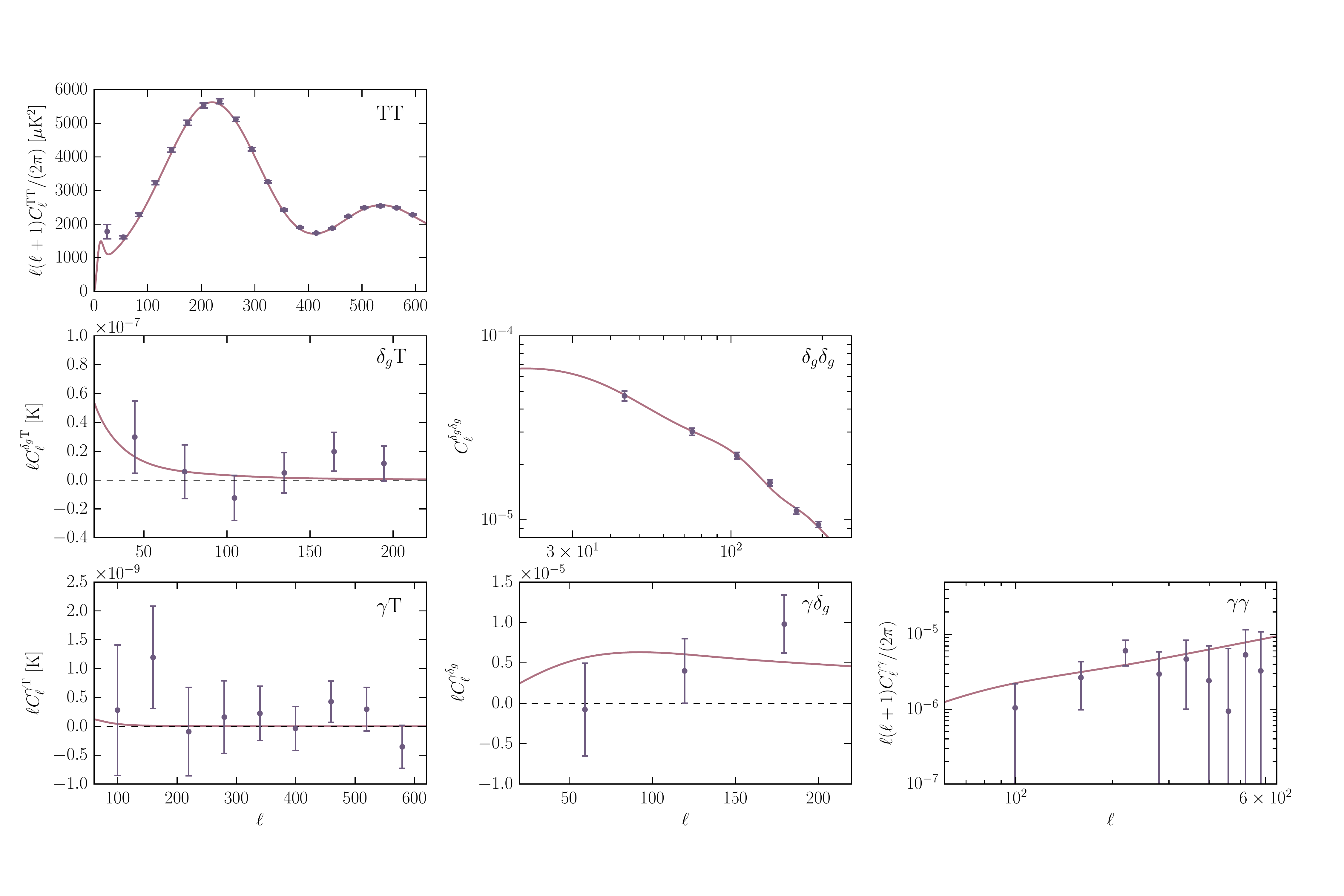}
\caption{Spherical harmonic power spectra for all probes used in the cosmological analysis. The top left panel shows the power spectrum of CMB anisotropies computed from the $\tt Commander$ CMB temperature map at resolution of $\tt NSIDE$ $= 2048$. The middle left panel shows the cross-power spectrum between CMB temperature anisotropies and galaxy overdensity computed from the systematics-reduced SDSS CMASS1-4 map and the $\tt{Commander}$ map at resolution $\tt NSIDE$ $= 1024$. The middle right panel shows the spherical harmonic power spectrum of the galaxy overdensity computed from the systematics-reduced SDSS CMASS1-4 map at $\tt NSIDE$ $= 1024$. The bottom left panel shows the spherical harmonic power spectrum between CMB temperature anisotropies and weak lensing shear measured from the $\tt{Commander}$ CMB map and the SDSS Stripe 82 weak lensing maps at resolution $\tt NSIDE$ $= 1024$. The bottom-middle panel shows the spherical harmonic power spectrum between galaxy overdensity and galaxy weak lensing shear computed from the systematics-reduced SDSS CMASS1-4 map and the SDSS Stripe 82 galaxy weak lensing shear map at resolution $\tt NSIDE$ $= 1024$. The bottom right panel shows the spherical harmonic power spectrum of cosmic shear E-modes computed from the SDSS Stripe 82 weak lensing shear maps. The angular multipole ranges and binning schemes for all power spectra are summarised in Table \ref{tab:clparams}. All power spectra are derived from the maps in Galactic coordinates. The solid lines show the theoretical predictions for the best-fit cosmological model determined from the joint analysis which is summarised in Tab.~\ref{tab:params}. The theoretical predictions have been convolved with the $\tt{PolSpice}$ kernels as described in Section \ref{sec:cls}. The error bars are derived from the Gaussian simulations described in Section \ref{sec:mockcov} and Appendix \ref{sec:corrmaps}.}
\label{fig:cls}
\end{figure*}

\subsection{\label{subsec:deltagcls} Galaxy clustering}

The galaxy overdensity maps described in Section \ref{sec:maps} are estimated from discrete galaxy tracers. Therefore, their spherical harmonic power spectrum receives contributions from the galaxy clustering signal and Poisson shot noise. To estimate the noise power spectrum, we resort to simulations. We generate noise maps by randomising the positions of all the galaxies in the sample inside the mask. Since this procedure removes all correlations between galaxy positions, the power spectra of these maps will give an estimate of the level of Poisson shot noise present in the data. In order to obtain a robust noise power spectrum, we generate 100 noise maps and estimate the noise power spectrum from the mean of these power spectra.

The spherical harmonic galaxy clustering power spectrum contains significant contributions from nonlinear structure formation at small angular scales. The effects of nonlinear galaxy bias are difficult to model and we therefore restrict our analysis to angular scales for which nonlinear corrections are small. We can estimate the significance of nonlinear effects by comparing the spherical harmonic galaxy clustering power spectrum computed using the nonlinear matter power spectrum as well as the linear matter power spectrum. Since galaxies are more clustered than dark matter this is likely to underestimate the effect. We find that the difference between the two reaches $5 \%$ of the power spectrum uncertainties and thus becomes mildly significant at around $\ell_{\mathrm{max}} \sim 250$. This difference is smaller than the difference derived in \citet{Ho:2012} and \citet{de-Putter:2012} which is likely due to the fact that we consider a single redshift bin and do not split the data into low and high redshifts. In order not to bias our results we choose $\ell_{\mathrm{max}} = 210$ which is comparable to the limit used in \citet{Ho:2012} and \citet{de-Putter:2012}. To determine the minimal angular multipole we follow \citet{Ho:2012}, who determined that the Limber approximation becomes accurate for scales larger than $\ell = 30$. 

The middle right panel in Fig.~\ref{fig:cls} shows the spherical harmonic galaxy clustering power spectrum computed from the systematics-corrected map in Galactic coordinates. In Appendix \ref{sec:cltests}, we compare the spherical harmonic power spectrum derived from the systematics-corrected maps in Galactic and equatorial coordinates. We find small differences at large angular scales, but the effect on the bandpowers considered in this analysis is negligible, as can be seen from Appendix \ref{sec:cltests} (Fig.~\ref{fig:eclgalcls}). To test the procedure for removing systematic uncertainties, we compare the spherical harmonic power spectra before and after correcting the maps for residual systematics. We find that the removal of systematics marginally reduces the clustering amplitude on large scales, which is expected since Galactic foregrounds exhibit significant large scale clustering. Small angular scales on the other hand, are mostly unaffected by the corrections applied. These results are shown in Appendix \ref{sec:cltests} (Fig.~\ref{fig:deltagcorrnocorrcls}). 

\subsection{\label{subsec:gammacls} Cosmic shear}

The power spectrum computed from the weak lensing shear maps described in Section \ref{subsec:ellipticitymap} contains contributions from both the cosmic shear signal and the shape noise of the galaxies, which is due to intrinsic galaxy ellipticities. In order to estimate the shape noise power spectrum we follow the same methodology as for galaxy clustering and resort to simulations. We generate noise-only maps by rotating the shears of all the galaxies in our sample by a random angle. This procedure removes spatial correlations between galaxy shapes. Since the weak lensing shear signal is at least an order of magnitude smaller than the intrinsic galaxy ellipticities, the power spectrum of the randomised map gives an estimate of the shape noise power spectrum. As for galaxy clustering, we compute 100 noise maps and estimate the shape noise power spectrum from the mean of these 100 noise power spectra.

For the cosmological analysis we choose broader multipole bins than for the CMB temperature anisotropies and galaxy clustering since the small sky fraction covered by SDSS Stripe 82 causes the cosmic shear power spectrum to be correlated across a significantly larger multipole range. The low and high $\ell$ limits are chosen to minimise demasking uncertainties and the impact of nonlinearities in the cosmic shear power spectrum.

The spherical harmonic power spectrum of the weak lensing shear E-mode is displayed in the bottom right panel of Fig.~\ref{fig:cls} and the B-mode power spectrum is shown in the Appendix (Fig.~\ref{fig:shearclsb}). We see that the E-mode power spectrum is intrinsically low as compared to the best-fit theory power spectrum. These results are similar to those derived by \citet{Lin:2012}, who found a low value of $\Omega_{\mathrm{m}}^{0.7}\sigma_{8}$ for Stripe 82 cosmic shear. As can be seen, we do not detect a significant B-mode signal. 

When comparing the weak lensing shear E-mode power spectra computed from the maps in Galactic and equatorial coordinates, we find discrepancies. These are mainly caused by the correction for additive bias in the weak lensing shears. As described in Appendix \ref{sec:cosmicshearanalysis}, the PSF-corrected galaxy shears are affected by an additive bias. Following \citet{Lin:2012}, we correct for this bias by subtracting the mean shear of each CCD camera column from the galaxy shears. This correction is performed in equatorial coordinates and ensures that the mean shear vanishes in this coordinate system. When the galaxy positions and shears are rotated from equatorial to Galactic coordinates, this ceases to be true. Therefore the correction for additive bias is coordinate-dependent and it is this effect that causes the main discrepancies between the measured power spectra. Further descriptions of the impact of the additive shear bias correction can be found in Appendix \ref{subsec:eclgal}.

The discrepancies between the cosmic shear power spectra measured from maps in Galactic and equatorial coordinates are still within the experimental uncertainties. We therefore choose to correct for the additive shear bias in equatorial coordinates, apply the rotation to the corrected shears and compute the cosmic shear power spectrum from the maps in Galactic coordinates. We note however, that these differences will become significant for surveys measuring cosmic shear with higher precision. It is therefore important to develop coordinate-independent methods for shear bias correction when performing a joint analysis of different cosmological probes.

\subsection{\label{subsec:tempcrossdeltagcls} CMB and galaxy overdensity cross-correlation}

To compute the spherical harmonic cross-power spectrum between CMB temperature anisotropies and the galaxy overdensity, we use the maps and masks described in Sections \ref{subsec:cmbmap} and \ref{subsec:galaxymap}. 

We generally have two possibilities to compute cross-correlations between two maps with different angular masks. We can either compute the cross-correlation by keeping the respective mask for each probe, or we can compute a combined mask, which is the union of all pixels masked in at least one of the maps. When testing both these cases on Gaussian simulations, we observed a better recovery of the input power spectra when applying the combined mask to both maps. We therefore mask both maps with the combined  mask, which covers a fraction of sky $f_{\mathrm{sky}} \sim 0.26$. 

The spherical harmonic cross-power spectrum between CMB temperature anisotropies and galaxy overdensity is shown in the middle left panel of Fig.~\ref{fig:cls}. We see that the ISW power spectrum is very noisy, which makes its detection significance small. Since the power spectrum uncertainties for the considered angular scales are mainly due to cosmic variance, we suspect that the low signal-to-noise is mainly due to the fraction of sky covered by the SDSS CMASS1-4 galaxies. Despite its low significance, we include the ISW power spectrum in our analysis, because we expect it to help break degeneracies between cosmological parameters. We check that the ISW power spectrum does not depend on the choice of foreground-reduced CMB map. We find that the results using the maps provided by the $\tt NILC$, $\tt SEVEM$ and $\tt SMICA$ algorithms are virtually the same, as illustrated in Appendix \ref{sec:cltests} (Fig.~\ref{fig:compplanckmaps}).

\subsection{\label{subsec:tempcrossgammacls} CMB and weak lensing shear cross-correlation}

We estimate the spherical harmonic cross-power spectrum between CMB temperature anisotropies and the weak lensing shear E-mode field from the maps and masks described in Sections \ref{subsec:cmbmap} and \ref{subsec:ellipticitymap}. Both maps are masked with the combination of the masks, which covers a fraction of sky $f_{\mathrm{sky}} \sim 0.0065$.

The bottom left panel in Fig.~\ref{fig:cls} shows the spherical harmonic power spectrum between CMB temperature anisotropies and the weak lensing shear E-mode field. As can be seen, the noise level is too high to allow for a detection of the ISW correlation between CMB temperature anisotropies and weak lensing shear. This is to be expected due to the small sky fraction covered by the SDSS Stripe 82 galaxies and the intrinsically low signal-to-noise of this cross-correlation. Nevertheless, we include the power spectrum into the joint analysis to provide an upper limit on the ISW from weak lensing. The measured power spectrum is unaffected by the choice of CMB mapmaking method, as illustrated in Fig.~\ref{fig:compplanckmaps} in Appendix \ref{sec:cltests}.

\subsection{\label{subsec:deltagcrossgammacls} Galaxy overdensity and weak lensing shear cross-correlation}

We compute the spherical harmonic cross-power spectrum between the galaxy overdensity and weak lensing shear E-mode field from the maps and masks described in Sections \ref{subsec:galaxymap} and \ref{subsec:ellipticitymap}. We mask both maps with the combination of the two masks. The combined mask covers a sky fraction $f_{\mathrm{sky}} \sim 0.0053$. 

The spherical harmonic cross-power spectrum between galaxy overdensity and weak lensing shear E-mode is shown in the bottom-middle panel of Fig.~\ref{fig:cls}. We see that the signal-to-noise of the power spectrum is low at the angular scales considered. This is probably due to the small sky fraction covered by Stripe 82 galaxies. We nevertheless include this cross-correlation in our analysis to serve as an upper limit. In Appendix \ref{sec:cltests} we show the comparison between the power spectra measured from the maps in Galactic and in equatorial coordinates. We find reasonable agreement between the two, even though the discrepancies are significantly enhanced compared to the effects on the galaxy overdensity power spectrum. As discussed in Section \ref{subsec:gammacls} this is probably due to the coordinate-dependence of the additive shear bias correction. 
 
\section{\label{sec:covariance} Covariance matrix}

In order to obtain cosmological constraints from a joint analysis of CMB temperature anisotropies, galaxy clustering and weak lensing we need to estimate the joint covariance matrix of these cosmological probes. In this work we assume all the fields to be Gaussian random fields, i.e. we assume the covariance between all probes to be Gaussian and neglect any non-Gaussian contribution. This is appropriate for the CMB temperature field as well as the galaxy overdensity field at the scales considered but it is only an approximation for the weak lensing shear field \cite{Sato:2009}. For example, for a survey with source redshifts $z_{s} = 0.6$, \citet{Sato:2009} found that neglecting non-Gaussian contributions leads to an underestimation of the diagonal terms in the cosmic shear covariance matrix by a factor of approximately 5 at multipoles $\ell \sim 600$. In our case the discrepancy may be more pronounced since our sample contains a significant number of galaxies with $z_{s} < 0.6$. On the other hand we will be less sensitive to the non-Gaussian nature of the covariance matrix since the covariance for our galaxy sample is dominated by shape noise especially at the highest multipoles considered. We therefore decide to leave the introduction of non-Gaussian covariance matrices to future work. 

In this work, we employ two different models for the joint Gaussian covariance matrix $C_{G}$: the first is a theoretical model and the second is based on simulations of correlated Gaussian realisations of the three cosmological probes. We use the theoretical covariance matrix to validate the covariance matrix obtained from the simulations.

\subsection{\label{sec:theorycov}Theoretical covariance estimate}

The covariance between cosmological spherical harmonic power spectra is composed of two parts: cosmic variance and noise. For spherical harmonic power spectra computed over the full sky, different $\ell$ modes are uncorrelated and the covariance matrix is diagonal. Partial sky coverage, i.e. $f_{\mathrm{sky}}<1$, has the effect to couple different $\ell$ modes and thus leads to a non-diagonal covariance matrix. This covariance becomes approximately diagonal if it is binned into approximately uncorrelated bandpowers of width $\Delta \ell$ \cite{Cabre:2007}. \citet{Cabre:2007} found the empirical relation $\Delta \ell f_{\mathrm{sky}} \sim 2$ to be a good approximation. In this case the covariance matrix between binned power spectra $C_{\ell}^{ij}$ and $C_{\ell'}^{i'j'}$ can be approximated as \cite{Hu:2004, Cabre:2007, Eifler:2014}
\begin{equation}
\begin{aligned}
\mathrm{Cov}_{G}(C_{\ell}^{ij}, C_{\ell'}^{i'j'}) = \langle \Delta C_{\ell}^{ij} \Delta C_{\ell'}^{i'j'} \rangle &\simeq  \\
\frac{\delta_{\ell \ell'}}{(2\ell+1)\Delta \ell f_{\mathrm{sky}}} \left [(C_{\ell}^{ii'} + N^{ii'})(C_{\ell}^{jj'} + N^{jj'})  \right. \\ 
+ \left. (C_{\ell}^{ij'} + N^{ij'})(C_{\ell}^{ji'} + N^{ji'})\right ],
\end{aligned}
\label{eq:theorycovmat}
\end{equation} 
where $i,\,j,\,i',\,j'$ denote different cosmological probes; in our case $i,\,j,\,i',\,j' \in \{\mathrm{T}, \delta_{g}, \gamma\}$. The quantities $N^{ij}$ are the noise power spectra of the different probes, which vanish unless $i = j$. 

Given a cosmological model and survey specifications such as fractional sky coverage and noise level, we can approximate $C_{G}$ using Eq.~\ref{eq:theorycovmat} for each block covariance matrix. We choose a hybrid approach: we adopt a cosmological model to calculate the signal power spectra whereas we approximate $N^{ij}$ with the measured noise power spectra used to remove the noise bias in the data as described in Section \ref{sec:cls}.

\subsection{\label{sec:mockcov} Covariance estimate from Gaussian simulations}

The theoretical covariance matrix estimate described above is expected to only yield accurate results for uncorrelated binned power spectra, since in this approximation the covariance matrix is fully diagonal. For this reason we also estimate the covariance matrix in an alternative way that does not rely on this approximation: we estimate an empirical covariance matrix from the sample variance of Gaussian simulations of the three cosmological probes. To this end, we simulate correlated realisations of both the two spin-0 fields, CMB temperature and galaxy overdensity, as well as the spin-2 weak lensing shear field. We follow the approach outlined in \citet{Giannantonio:2008} for simulating correlated maps of spin-0 fields and we make use of the polarisation version of the HEALPix routine $\tt synfast$ to additionally simulate correlated maps of the spin-2 field. We estimate noise maps from the data and add these to the correlated signal maps. The details of the algorithm are outlined in Appendix \ref{sec:corrmaps}.

In order to compute the power spectrum covariance matrix, we apply the masks used on the data to the simulated maps and calculate both the auto- and the cross-power spectra of all the probes using the same methodology and $\tt{PolSpice}$ settings as described in Section \ref{sec:cls}. We generate $N_{\mathrm{sim}}$ random realisations and estimate the covariance matrix as 
\begin{equation}
\begin{aligned}
\mathrm{Cov}_{G}(C_{\ell}^{ij}, C_{\ell'}^{i'j'}) = \frac{1}{N_{\mathrm{sim}}-1} \sum_{k=1}^{N_{\mathrm{sim}}} \left[C_{k}^{ij}(\ell) - \bar{C}_{k}^{ij}(\ell)\right] \\
\times \left[C_{k}^{i'j'}(\ell') - \bar{C}_{k}^{i'j'}(\ell')\right],
\end{aligned}
\end{equation}
where $\bar{C}_{k}^{ij}(\ell)$ denotes the mean over all realisations.

The accuracy of the sample covariance estimate depends on the number of simulations. As described in \citet{Cabre:2007}, $N_{\mathrm{sim}} = 1000$ achieves better than $5 \%$ accuracy for estimating the covariance matrix for the ISW effect from Gaussian simulations. We therefore follow \citet{Cabre:2007} and compute the covariance matrix from the sample variance of $N_{\mathrm{sim}} = 1000$ Gaussian realisations of the 4 maps or 6 spherical harmonic power spectra respectively. 

\begin{figure}
\includegraphics[scale=0.32]{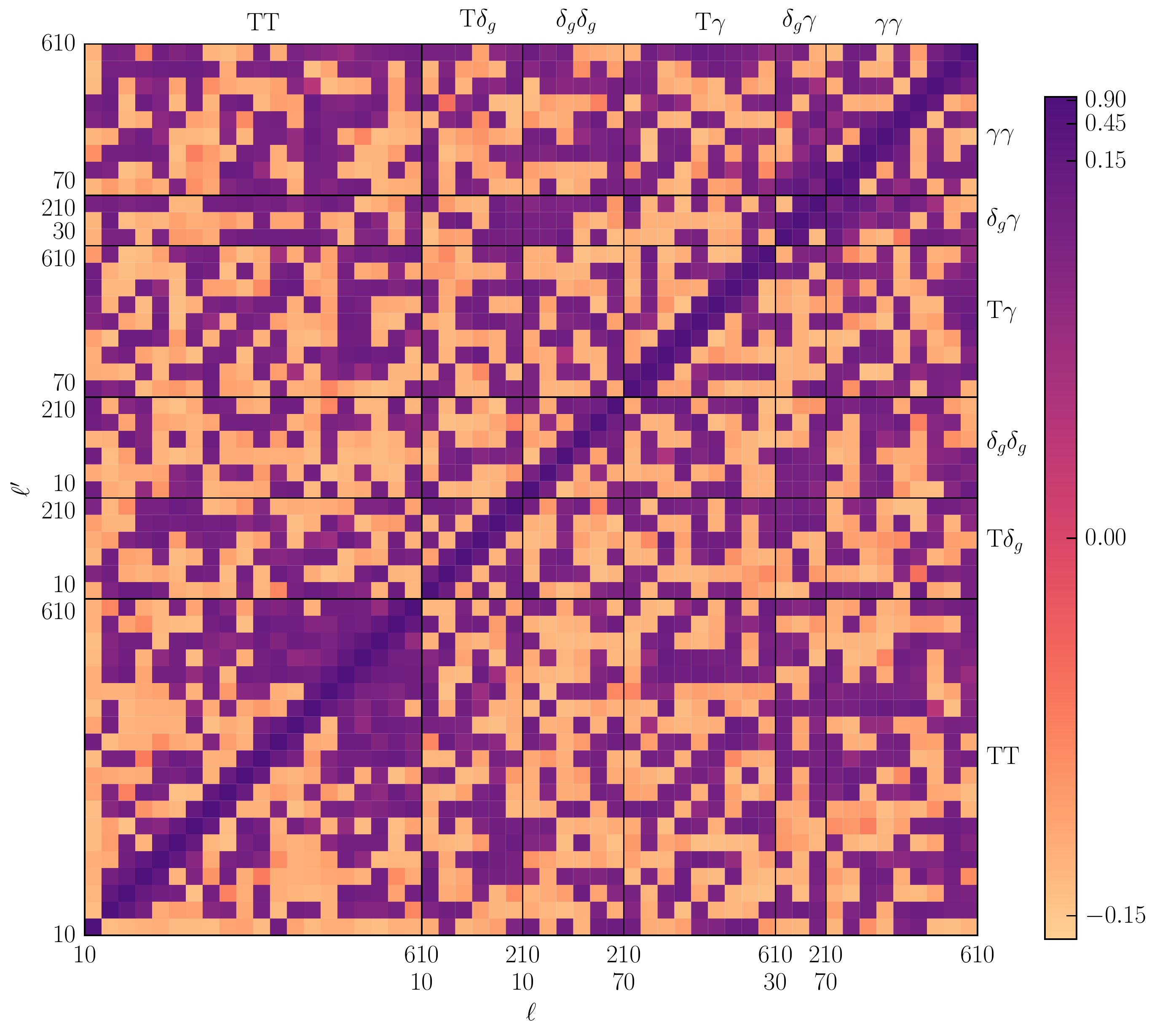}
\caption{Correlation matrix for the spherical harmonic power spectra derived from the sample variance of the Gaussian simulations. The binning scheme and angular multipole range for each probe follow those outlined in Tab.~\ref{tab:clparams}.}
\label{fig:mockcovfull}
\end{figure}

The correlation matrix for the spherical harmonic power spectra derived from the Gaussian simulations for binning schemes and angular multipole ranges described in Section \ref{sec:cls} is shown in Fig.~\ref{fig:mockcovfull}. We see that the survey masks lead to significant correlations between bandpowers. 

\section{\label{sec:results} Cosmological constraints}

Each of the power spectra presented in Section \ref{sec:cls} carries cosmological information with probe-specific sensitivities and degeneracies. An integrated combination of these cosmological probes therefore helps break these parameter degeneracies. It further provides robust cosmological constraints since it is derived from a joint fit to the auto- as well as cross-correlations of three cosmological probes. 

In order to derive cosmological constraints from a joint fit to the six spherical harmonic power spectra discussed in Section \ref{sec:cls}, we assume the joint likelihood to be Gaussian, i.e.
\begin{multline}
\mathscr{L}(D \vert \theta) = \frac{1}{[(2\pi)^{d}\det{C_{G}}]^{\sfrac{1}{2}}} \\
\times e^{-\frac{1}{2}(\mathbf{C}^{\mathrm{obs}}_{\ell}-\mathbf{C}^{\mathrm{theor}}_{\ell})^{\mathrm{T}}C^{-1}_{G}(\mathbf{C}^{\mathrm{obs}}_{\ell}-\mathbf{C}^{\mathrm{theor}}_{\ell})},
\label{eq:likelihood}
\end{multline}
where $C_{G}$ denotes the Gaussian covariance matrix. $\mathbf{C}^{\mathrm{theor}}_{\ell}$ denotes the theoretical prediction for the spherical harmonic power spectrum vector of dimension $d$ and $\mathbf{C}^{\mathrm{obs}}_{\ell}$ is the observed power spectrum vector, defined as 
\begin{equation}
\mathbf{C}^{\mathrm{obs}}_{\ell} = \begin{pmatrix}
C^{\mathrm{TT}}_{\ell} & C_{\ell}^{\delta_{g}\mathrm{T}} & C_{\ell}^{\delta_{g} \delta_{g}} & C^{\gamma\mathrm{T}}_{\ell} & C^{\gamma \delta_{g}}_{\ell} & C^{\gamma \gamma}_{\ell}
\end{pmatrix}_{\mathrm{obs}}.
\label{eq:psvector}
\end{equation} 
A Gaussian likelihood is a justified assumption for both the CMB temperature anisotropy and galaxy clustering power spectra due to the central limit theorem. Since the weak lensing shear power spectrum receives significant contribution from non-linear structure formation, its likelihood will deviate from being purely Gaussian \cite{Hartlap:2009}. It has been shown however, that a Gaussian likelihood is a sensible approximation, especially when CMB data is added to weak lensing \cite{Sato:2010}. In our first implementation we will thus assume both a joint Gaussian likelihood and Gaussian single probe likelihoods.

We estimate the covariance matrix using both methods outlined in Section \ref{sec:covariance}. In both cases we compute the covariance for a $\Lambda$CDM cosmological model, which we keep fixed in the joint fit. Note that the covariance matrices depend on the cosmological model and should therefore vary in the fitting procedure \cite{Eifler:2009}. Following standard practice, (e.g. \cite{DES-Collaboration:2015}), we approximate the covariance matrix to be constant and compute it for a $\Lambda$CDM cosmological model with parameter values $\{h,\, \Omega_{\mathrm{m}}, \,\Omega_{\mathrm{b}}, \, n_{\mathrm{s}}, \,\sigma_{8}, \,\tau_{\mathrm{reion}}, \,T_{\mathrm{CMB}}\} = \{0.7, \,0.3, \,0.049, \,1.0, \,0.88, \,0.078, \,2.275 \,\mathrm{K}\}$, where $h$ is the dimensionless Hubble parameter, $\Omega_{\mathrm{m}}$ is the fractional matter density today, $\Omega_{\mathrm{b}}$ is the fractional baryon density today, $n_{\mathrm{s}}$ denotes the scalar spectral index, $\sigma_{8}$ is the r.m.s. of matter fluctuations in spheres of comoving radius $8 \,h^{-1}$ Mpc and $\tau_{\mathrm{reion}}$ denotes the optical depth to reionisation. We further set the linear, redshift-independent galaxy bias parameter to $b=2$. To obtain an unbiased estimate of the inverse of the covariance matrix derived from the Gaussian simulations, we apply the correction derived in \citet{Kaufman:1967}, \citet{Hartlap:2007} and \citet{Anderson:2003}, i.e. we multiply the inverse covariance matrix by $(N_{\mathrm{sim}}-d-2)/(N_{\mathrm{sim}}-1)$. The theoretical covariance matrix estimate does not suffer from this bias and is thus left unchanged. 

From the likelihood given in Eq.~\ref{eq:likelihood}, we derive constraints in the framework of a flat $\Lambda$CDM cosmological model, where our fiducial model includes one massive neutrino eigenstate of mass $0.06$ eV as in \cite{Planck-Collaboration:2015ae}. Our parameter set consists of the six $\Lambda$CDM parameters $\{h,\, \Omega_{\mathrm{m}}, \,\Omega_{\mathrm{b}}, \,n_{\mathrm{s}}, \,\sigma_{8}, \,\tau_{\mathrm{reion}}\}$. We further marginalise over two additional parameters: a redshift independent, linear galaxy bias parameter $b$ and a multiplicative bias parameter $m$ for the weak lensing shear. The multiplicative bias parametrises unaccounted calibration uncertainties affecting the weak lensing shear estimator $\hat{\boldsymbol{\gamma}}$ and is defined as \cite{Heymans:2006}
\begin{equation}
\hat{\boldsymbol{\gamma}} = (1+m)\boldsymbol{\gamma}.
\end{equation}
We note that we do not include additional nuisance parameters such as additive weak lensing shear bias, stochastic and scale-dependent galaxy bias \cite{Tegmark:1998, Pen:1998, Dekel:1999}, photometric redshift uncertainties, intrinsic galaxy alignments (for reviews, see e.g. \cite{Troxel:2015, Joachimi:2015}) or parameters describing the effect of unresolved point sources on the CMB temperature anisotropy power spectrum \cite{Planck-Collaboration:2014af}. In this present work we restrict the analysis to angular scales where these effects are expected to be subdominant.

We sample the parameter space with a Monte Carlo Markov Chain (MCMC) using $\tt CosmoHammer$ \cite{Akeret:2012}. The parameters sampled are summarised in Table \ref{tab:params} along with their priors. We choose flat, uniform priors for all parameters except for $\tau_{\mathrm{reion}}$ and $m$. The optical depth to reionisation can only be constrained with CMB polarisation data. Since we do not include CMB polarisation in this analysis, we apply a Gaussian prior with $\mu = 0.089$ and $\sigma = 0.02$ on $\tau_{\mathrm{reion}}$. This corresponds to a WMAP9 \cite{Hinshaw:2013} prior with increased variance to accommodate the Planck 2015 results \cite{Planck-Collaboration:2015ae}. We further apply a Gaussian prior on the multiplicative bias $m$ with mean $\mu=0$ and $\sigma=0.1$. This is motivated by \citet{Hirata:2003}, who found the multiplicative bias for the linear PSF correction method to lie in the range $m \in [-0.08, 0.13]$ for the sample considered in this analysis. 

\begin{table*}
\caption{Parameters varied in the MCMC with their respective priors and posterior means. The uncertainties denote the $68 \%$ c.l..} \label{tab:params}
\begin{center}
\begin{ruledtabular}
\begin{tabular}{ccc}
Parameter & Prior & Posterior mean\\ \hline \Tstrut                             
$h$ & flat $\in [0.2, \,1.2]$ & $0.699 \pm 0.018$ \\ 
$\Omega_{\mathrm{m}}$ & flat $\in [0.1, \,0.7]$ & $0.278 \substack{+0.019 \\ -0.020}$ \\
$\Omega_{\mathrm{b}}$ & flat $\in [0.01, \,0.09]$ & $0.0455 \pm 0.0018$ \\
$n_{\mathrm{s}}$ & flat $\in [0.1, \,1.8]$ & $0.975 \substack{+0.019 \\ -0.018}$ \\
$\sigma_{8}$ & flat $\in [0.4, \,1.5]$ & $0.799 \pm 0.029$ \\ \\
$\tau_{\mathrm{reion}}$ & Gaussian with $\mu = 0.089$, $\sigma = 0.02$\footnote{This corresponds to a WMAP9 \cite{Hinshaw:2013} prior with increased variance to accommodate the Planck results.} & $0.0792  \pm 0.0196$ \\
$b$ & flat $\in [1., \,3.]$ & $2.13 \pm 0.06$ \\
$m$ & Gaussian with $\mu = 0.0$, $\sigma = 0.1$ & $-0.142 \substack{+0.080 \\ -0.081}$
\end{tabular}
\end{ruledtabular}
\end{center}
\end{table*} 

In our fiducial configuration presented below we use the covariance matrix derived from the Gaussian simulations as described in Section \ref{sec:mockcov}. We find that this choice does not influence our results since the constraints derived using the theoretical covariance are consistent. In order to further assess the impact of a cosmology-dependent covariance matrix, we perform the equivalent analysis using a covariance matrix computed with a cosmological model with $\sim 5 \%$ lower $\sigma_{8}$. We find that the derived parameter values change by at most $0.5\sigma$. The width of the contours is only marginally changed. 

In addition to the joint analysis, we also derive parameter constraints from separate analyses of the three auto-power spectra $C^{\mathrm{TT}}_{\ell}, C_{\ell}^{\delta_{g} \delta_{g}}$ and $C^{\gamma \gamma}_{\ell}$. In all three cases we assume a Gaussian likelihood as in Eq.~\ref{eq:likelihood} and derive constraints on the base $\Lambda$CDM parameters $\{h,\, \Omega_{\mathrm{m}}, \,\Omega_{\mathrm{b}}, \,n_{\mathrm{s}}, \,\sigma_{8}\}$ as well as additional parameters constrained by each probe. These are $\tau_{\mathrm{reion}}$ for the CMB temperature anisotropies, $b$ for galaxy clustering and $m$ for the cosmic shear.

\begin{figure*}
\includegraphics[scale=0.6]{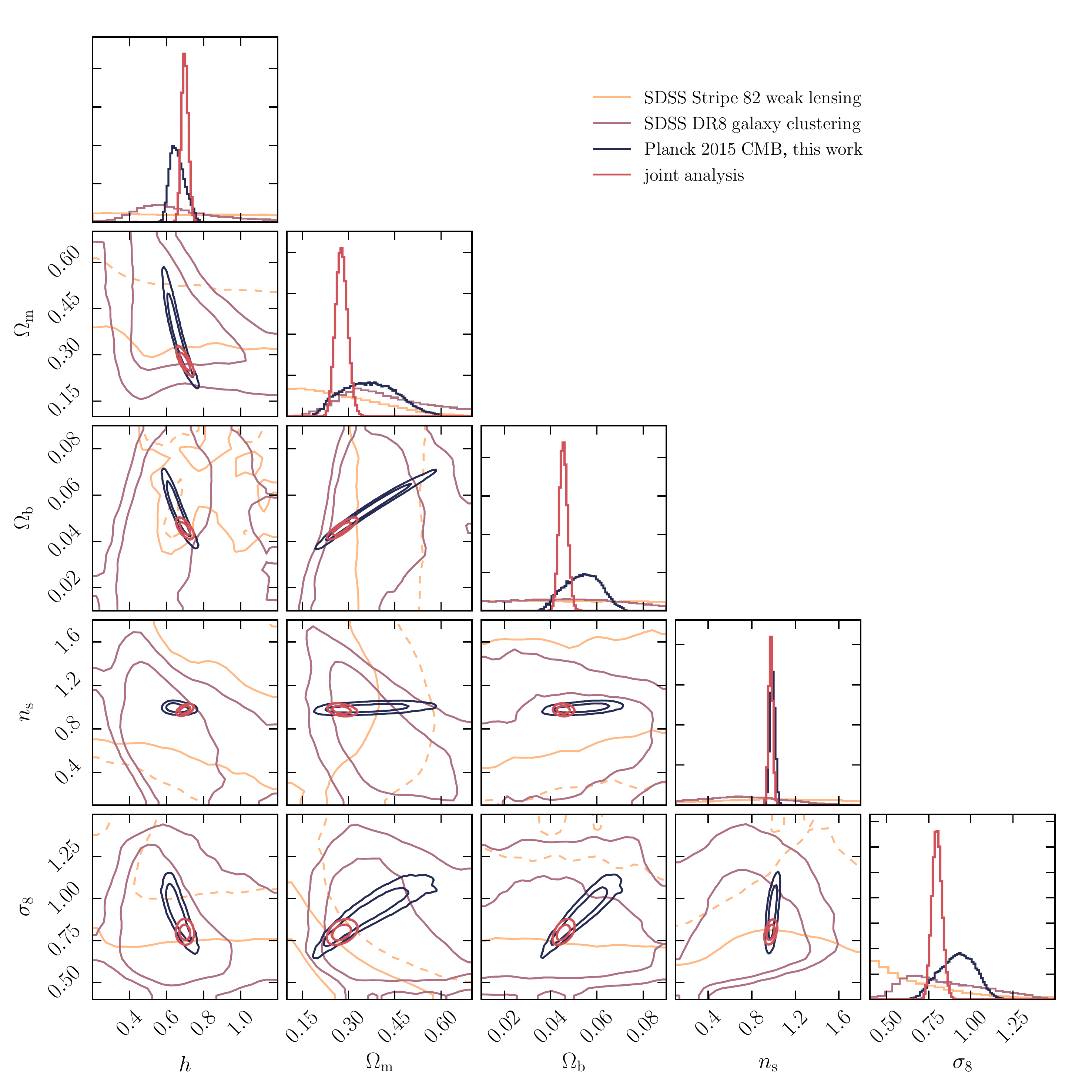}
\caption{Cosmological parameter constraints derived from the joint analysis, marginalised over $\tau_{\mathrm{reion}}$, $b$ and $m$ and from the single probes. The single probe constraints have been marginalised over the respective nuisance parameters i.e. $\tau_{\mathrm{reion}}$ for the CMB temperature anisotropies, $b$ for galaxy clustering and $m$ for the weak lensing shear. In each case the inner (outer) contour shows the $68 \%$ c.l. ($95 \%$ c.l.). For clarity the cosmic shear $68 \%$ c.l. are solid while the $95 \%$ c.l. are dashed. }
\label{fig:constraintssinglevsjoint}
\end{figure*}

\begin{figure*}
\includegraphics[scale=0.5]{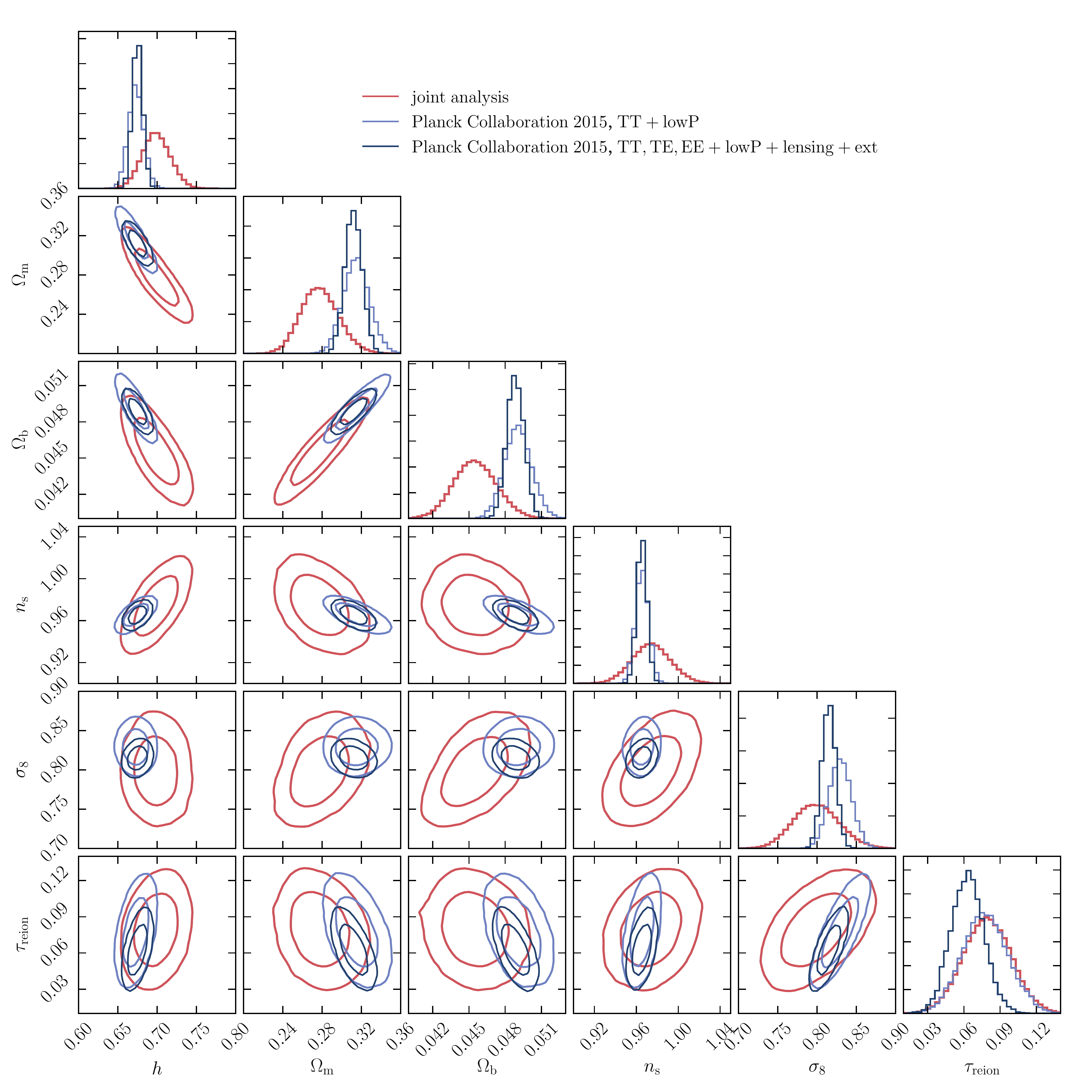}
\caption{Comparison between the parameter constraints derived from the joint analysis, marginalised over $b$ and $m$ and the constraints from \citet{Planck-Collaboration:2015ae} using only CMB data (TT+lowP) or adding external data (TT,TE,EE+lowP+lensing+ext). The Planck constraints are marginalised over all nuisance parameters. In each case the inner (outer) contour shows the $68 \%$ c.l. ($95 \%$ c.l.).}
\label{fig:constraintscompplanck}
\end{figure*}

Fig.~\ref{fig:constraintssinglevsjoint} shows the constraints on the $\Lambda$CDM parameters $\{h,\, \Omega_{\mathrm{m}}, \,\Omega_{\mathrm{b}}, \,n_{\mathrm{s}}, \,\sigma_{8}\}$ derived from the joint analysis using the spherical harmonic power spectrum vector and likelihood defined in Equations \ref{eq:likelihood} and \ref{eq:psvector}. These have been marginalised over $\tau_{\mathrm{reion}}$, $b$ and $m$. Also shown are the constraints derived from separate analyses of the three auto-power spectra $C^{\mathrm{TT}}_{\ell}, C_{\ell}^{\delta_{g} \delta_{g}}$ and $C^{\gamma \gamma}_{\ell}$, each of them marginalised over the respective nuisance parameter. As expected, we find that the constraints derived from the CMB anisotropies are the strongest, followed by the galaxy clustering and cosmic shear constraints, which both constrain the full $\Lambda$CDM model rather weakly. The constraints from the CMB temperature anisotropies are broader and have central values which differ from those derived in \citet{Planck-Collaboration:2015ae}. The reason for these discrepancies is the limited angular multipole range $\ell \in [10, \,610]$ employed in the CMB temperature analysis. This causes the CMB posterior to become broader, asymmetric and results in a shift of the parameter means. We have verified that the Planck likelihood and our analysis give consistent results when the latter is restricted to a similar $\ell$-range. If on the other hand, we increase the high multipole limit to $\ell_{\mathrm{max}} = 1000$, we find significant differences between our analysis and the Planck likelihood. We therefore choose to be conservative and use $\ell_{\mathrm{max}} = 610$ throughout this work. Comparing the single probe constraints to one another we see that they agree reasonably well, the only slight discrepancy being the low value of both $\Omega_{\mathrm{m}}$ and $\sigma_{8}$ derived from the cosmic shear analysis. This is similar to the results derived in \citet{Lin:2012} even though the values for $\Omega_{\mathrm{m}}$ and $\sigma_{8}$ are even lower in our analysis. However, care must be taken since the amplitude of the cosmic shear auto-power spectrum appears to have a small dependence on the choice of the coordinate system as discussed in Appendix \ref{sec:cltests}.

The potential of the joint analysis emerges when the three auto-power spectra are combined together with their three cross-power spectra. Due to the complementarity of the different probes the constraints tighten and the allowed parameter space volume is significantly reduced. This is especially true in our case, since the constraints from CMB temperature anisotropies are broadened due to the restricted multipole range that we employed. Including more CMB data would significantly reduce the impact of adding additional cosmological probes. The numerical values of the best-fit parameters and their $68 \%$ confidence limits (c.l.) derived from the joint analysis are given in Tab.~\ref{tab:params}.

Fig.~\ref{fig:constraintscompplanck} compares the constraints derived from the joint analysis to the constraints derived by the Planck Collaboration \cite{Planck-Collaboration:2015ae}. We show two versions of the Planck constraints: the constraints derived from the combination of CMB temperature anisotropies with the Planck low-$\ell$ polarisation likelihood (TT+lowP) and the ones derived from a combination of the latter with the Planck polarisation power spectra, CMB lensing and external data sets (TT,TE,EE+lowP+lensing+BAO+JLA+$H_{0}$). We see that the joint analysis prefers slightly lower values of the parameters $\Omega_{\mathrm{m}}$ and $\Omega_{\mathrm{b}}$ and a higher Hubble parameter $h$, but these differences are not significant. Despite this fact we find sensible overall agreement between the constraints derived in this work with both versions of the Planck constraints. While the constraints we derived in this analysis are broadened by the restricted multipole range we used, the results already demonstrate the power of integrated probe combination: the complementarity of different cosmological probes and their cross-correlations allows us to obtain reasonable constraints. 

The measured power spectra together with the theoretical predictions for the best-fitting cosmological model derived from the joint analysis are shown in Fig.~\ref{fig:cls}. The best-fit cosmology provides a rather good fit to all power spectra except $C^{\gamma \delta_{g}}_{\ell}$ and $C^{\gamma \gamma}_{\ell}$, whose measured values are generally lower than our best-fit model. This is mainly due to the assumed Gaussian prior on the multiplicative shear bias $m$, which does not allow for more negative values of $m$ as would be preferred by the data. If we relax the prior to a Gaussian with standard deviation $\sigma=0.2$, we find a best-fit value for the multiplicative bias parameter of $m = -0.276 \pm 0.108$. This results in an improved fit to both $C^{\gamma\delta_{g}}_{\ell}$ and $C^{\gamma \gamma}_{\ell}$, but is in tension with the values derived for the multiplicative bias by \citet{Hirata:2003}. We therefore find evidence for a slight tension between CMB temperature anisotropy data and weak gravitational lensing, as already seen by e.g. \cite{MacCrann:2015, Grandis:2016}.

\section{\label{sec:conclusions} Conclusions}

To further constrain our cosmological model and gain more information about the dark sector, it will be essential to combine the constraining power of different cosmological probes. This work presents a first implementation of an integrated approach to combine cosmological probes into a common framework at the map level. In our first implementation we combine CMB temperature anisotropies, galaxy clustering and weak lensing shear. We use CMB data from Planck 2015 \cite{Planck-Collaboration:2015af}, photometric galaxy data from the SDSS DR8 \cite{Aihara:2011} and weak lensing data from SDSS Stripe 82 \cite{Annis:2014}.
We take into account both the information contained in the separate maps as well as the information contained in the cross-correlation between the maps by measuring their spherical harmonic power spectra. This leads to a power spectrum matrix with associated covariance, which combines CMB temperature anisotropies, galaxy clustering, cosmic shear, galaxy-galaxy lensing and the ISW \cite{Sachs:1967} effect with galaxy and weak lensing shear tracers.

From the power spectrum matrix we derive constraints in the framework of a $\Lambda$CDM cosmological model assuming both a Gaussian covariance as well as a Gaussian likelihood. We find that the constraints derived from the combination of all probes are significantly tightened compared to the constraints derived from each of the three separate auto-power spectra. This is due to the complementary information carried by different cosmological probes. We further compare these constraints to existing ones derived by the Planck collaboration and find reasonable agreement, even though the joint analysis slightly prefers lower values of both $\Omega_{\mathrm{m}}$ and $\Omega_{\mathrm{b}}$ and a higher Hubble parameter $h$. For a joint analysis of three cosmological probes, the constraints derived are still relatively weak, which is mainly due to our conservative cuts in angular scales. Nevertheless this analysis already demonstrates the potential of integrated probe combination: the complementarity of different data sets, that alone yield rather weak constraints on the full $\Lambda$CDM parameter space, allows us to obtain robust constraints which are significantly tighter than those obtained from probes taken individually. In addition, our analysis reveals challenges intrinsic to probe combination. Examples are the need for foreground-correction at the map as opposed to the power spectrum level and the need for coordinate-independent bias corrections.

In this first implementation we have made simplifying assumptions. We assume a Gaussian covariance matrix for all cosmological probes considered. This is justified for the CMB temperature anisotropies and the galaxy overdensity at large scales. The galaxy shears on the other hand exhibit non-linearities already at large scales and their covariance therefore receives significant non-Gaussian contributions \cite{Sato:2009}. Furthermore, we do not take into account the cosmology-dependence of the covariance matrix \cite{Eifler:2009}. In addition we only include systematic uncertainties from a potential multiplicative bias in the weak lensing shear measurement and neglect effects from other sources. Finally we also used the Limber approximation for the theoretical predictions. We leave these extensions to future work but we do not expect them to have a significant impact on our results since we restrict the analysis to scales where the above effects are minimised. 

In order to fully exploit the wealth of cosmological information contained in upcoming surveys, it will be essential to investigate ways in which to combine these experiments. It will be thus interesting to extend the framework presented here to include additional cosmological probes, 3-dimensional tomographic information and tests of cosmological models beyond $\Lambda$CDM.

\begin{acknowledgments}

We thank Eric Hivon for his valuable help with $\tt PolSpice$. We also thank Chihway Chang for careful reading of the manuscript and helpful comments and Sebastian Seehars as well as Joel Akeret for helpful discussions. We further thank the referee for comments and suggestions that have improved this paper. This work was in part supported by the Swiss National Science Foundation (grant number 200021 143906).

Some of the results in this paper have been derived using the HEALPix (\citet{Gorski:2005}) package. Icons used in Fig.~\ref{fig:framework} made by Freepik from $\tt{www.flaticon.com}$. The colour palettes employed in this work are taken from $\tt{http://colorpalettes.net}$ and $\tt{http://flatuicolors.com}$. We further acknowledge the use of the colour map provided by \cite{Planck-Collaboration:2015af}. The contour plots have been created using $\tt{corner.py}$ \cite{ForemanMackey:2016}.  

Funding for the SDSS and SDSS-II has been provided by the Alfred P. Sloan Foundation, the Participating Institutions, the National Science Foundation, the U.S. Department of Energy, the National Aeronautics and Space Administration, the Japanese Monbukagakusho, the Max Planck Society, and the Higher Education Funding Council for England. The SDSS Web Site is http://www.sdss.org/.

The SDSS is managed by the Astrophysical Research Consortium for the Participating Institutions. The Participating Institutions are the American Museum of Natural History, Astrophysical Institute Potsdam, University of Basel, University of Cambridge, Case Western Reserve University, University of Chicago, Drexel University, Fermilab, the Institute for Advanced Study, the Japan Participation Group, Johns Hopkins University, the Joint Institute for Nuclear Astrophysics, the Kavli Institute for Particle Astrophysics and Cosmology, the Korean Scientist Group, the Chinese Academy of Sciences (LAMOST), Los Alamos National Laboratory, the Max-Planck-Institute for Astronomy (MPIA), the Max-Planck-Institute for Astrophysics (MPA), New Mexico State University, Ohio State University, University of Pittsburgh, University of Portsmouth, Princeton University, the United States Naval Observatory, and the University of Washington.

Funding for SDSS-III has been provided by the Alfred P. Sloan Foundation, the Participating Institutions, the National Science Foundation, and the U.S. Department of Energy Office of Science. The SDSS-III web site is http://www.sdss3.org/.

SDSS-III is managed by the Astrophysical Research Consortium for the Participating Institutions of the SDSS-III Collaboration including the University of Arizona, the Brazilian Participation Group, Brookhaven National Laboratory, Carnegie Mellon University, University of Florida, the French Participation Group, the German Participation Group, Harvard University, the Instituto de Astrofisica de Canarias, the Michigan State/Notre Dame/JINA Participation Group, Johns Hopkins University, Lawrence Berkeley National Laboratory, Max Planck Institute for Astrophysics, Max Planck Institute for Extraterrestrial Physics, New Mexico State University, New York University, Ohio State University, Pennsylvania State University, University of Portsmouth, Princeton University, the Spanish Participation Group, University of Tokyo, University of Utah, Vanderbilt University, University of Virginia, University of Washington, and Yale University.

Based on observations obtained with Planck (http://www.esa.int/Planck), an ESA science mission with instruments and contributions directly funded by ESA Member States, NASA, and Canada.

\end{acknowledgments}

\appendix

\section{\label{sec:shearisw}Theoretical prediction for CMB and weak lensing shear cross-correlation}

The CMB temperature anisotropies are correlated with the weak lensing shear due to the ISW effect. The anisotropies in the temperature field generated by time-varying gravitational potentials $\Phi$ are given by (see e.g. \cite{Padmanabhan:2005}):
\begin{equation}
\Delta T_{\mathrm{ISW}}(\boldsymbol{\theta}) = T_{\mathrm{CMB}} \,\delta T_{\mathrm{ISW}} = 2 \,T_{\mathrm{CMB}} \,\int_{\eta_{r}}^{\eta_{0}} \mathrm{d}\eta \frac{\partial \Phi}{\partial \eta},
\end{equation}
where $\eta_{0}$ denotes the conformal time today and $\eta_{r}$ is the conformal time at recombination. Note that we follow the conventions for the gravitational potential $\Phi$ as in \citet{Bartelmann:2010}.
These anisotropies can be decomposed into spherical harmonics with multipole coefficients
\begin{multline}
\Delta T_{\mathrm{ISW}, \ell m} = 4\pi i^{\ell} 2 \,T_{\mathrm{CMB}} \,\int_{\eta_{r}}^{\eta_{0}} \mathrm{d}\eta \\
\times \int \frac{\mathrm{d}^{3}\mathbf{k}}{(2\pi)^{3}} \frac{\mathrm{d}}{\mathrm{d} \eta}[\Phi(\mathbf{k}, z)] j_{\ell}\boldsymbol{(}k\chi(z)\boldsymbol{)} Y_{\ell m}^{*}(\boldsymbol{\theta}_{k}).
\end{multline}
The multipole coefficients of the weak lensing shear E-modes can be expressed through the lensing potential $\psi$ and are given by \cite{Bartelmann:2010}
\begin{multline}
a_{\mathrm{E}, \ell m} =-\frac{1}{2} \sqrt{\frac{(\ell+2)!}{(\ell-2)!}} \psi_{\ell, m} = - \sqrt{\frac{(\ell+2)!}{(\ell-2)!}} 4\pi i^{\ell} \\ 
\times \int \mathrm{d}\chi \,g(\chi) \int \frac{\mathrm{d}^{3}\mathbf{k}'}{(2\pi)^{3}} \Phi(\mathbf{k}', z) j_{\ell}\boldsymbol{(}k'\chi(z)\boldsymbol{)} Y_{\ell m}^{*}(\boldsymbol{\theta}_{k'}) ,
\end{multline}
where
\begin{equation}
g(\chi) = \frac{1}{\chi(z)} \int_{\chi(z)}^{\chi_{h}}\mathrm{d}z' \frac{\chi(z')-\chi(z)}{\chi(z')} n(z').
\end{equation}
The spherical harmonic power spectrum $C_{\ell}^{\gamma\mathrm{T}}$ between CMB temperature anisotropies and the weak lensing shear is defined as
\begin{equation}
\langle \Delta T_{\mathrm{ISW}, \ell m} a^{*}_{\mathrm{E}, \ell' m'} \rangle = C_{\ell}^{\gamma\mathrm{T}} \delta_{\ell \ell'}\delta_{m m'}.
\end{equation}
Expressing the integrals in terms of redshift and interchanging the integration boundaries gives
\begin{widetext}
\begin{multline}
\langle \Delta T_{\mathrm{ISW}, \ell m} a^{*}_{\mathrm{E}, \ell' m'} \rangle = (4\pi)^{2} \sqrt{\frac{(\ell+2)!}{(\ell-2)!}} 2 \,T_{\mathrm{CMB}} \langle \int_{0}^{z_{*}} \mathrm{d}z \int \frac{\mathrm{d}^{3}\mathbf{k}}{(2\pi)^{3}} \frac{\mathrm{d}}{\mathrm{d}z}[D(z)(1+z)]\Phi(\mathbf{k}, z=0) j_{\ell}\boldsymbol{(}k\chi(z)\boldsymbol{)} Y_{\ell m}^{*}(\boldsymbol{\theta}_{k}) \\
 \times \int \mathrm{d}z' \frac{c}{H(z')} g\boldsymbol{\left(}\chi(z')\boldsymbol{\right)} \int \frac{\mathrm{d}^{3}\mathbf{k}'}{(2\pi)^{3}} \Phi(\mathbf{k}', z')j_{\ell'}\boldsymbol{(}k'\chi(z')\boldsymbol{)} Y_{\ell' m'}(\boldsymbol{\theta}_{k'}) \rangle,
\label{eq:clsiswshear1}
\end{multline}
\end{widetext}
where $z_{*}$ denotes the redshift at recombination. In order to derive Eq.~\ref{eq:clsiswshear1} we have used that in linear perturbation theory the time- and scale-dependence of the gravitational potentials can be separated i.e.:
\begin{equation}
\Phi(k, z) = \Phi(k, z=0) D(z)(1+z),
\end{equation}
where $D(z)$ denotes the linear growth factor.
We further have that 
\begin{equation}
\langle \Phi(\mathbf{k}, z=0) \Phi(\mathbf{k}', z'=0) \rangle = (2\pi)^{3}\, P^{\mathrm{lin}}_{\Phi \Phi}(k, z=0)\, \delta(\mathbf{k}-\mathbf{k}'),
\end{equation}
and therefore Eq.~\ref{eq:clsiswshear1} reduces to 
\begin{widetext}
\begin{multline}
C_{\ell}^{\gamma\mathrm{T}} = (4\pi)^{2} \sqrt{\frac{(\ell+2)!}{(\ell-2)!}} \, 2 \,T_{\mathrm{CMB}} \int_{0}^{z_{*}} \mathrm{d}z \int \frac{k^{2}\mathrm{d}k}{(2\pi)^{3}} \,\frac{\mathrm{d}}{\mathrm{d}z}[D(z)(1+z)] \\
 \times \int \mathrm{d}z' \frac{c}{H(z')} \,g\boldsymbol{\left(}\chi(z')\boldsymbol{\right)} \,D(z')(1+z') \,P^{\mathrm{lin}}_{\Phi \Phi}(k, z=0)\, j_{\ell}\boldsymbol{(}k\chi(z)\boldsymbol{)}\, j_{\ell}\boldsymbol{(}k\chi(z')\boldsymbol{)}.
\label{eq:clsiswshear2}
\end{multline}
\end{widetext}
Eq.~\ref{eq:clsiswshear2} is the exact expression for the spherical harmonic cross-power spectrum between CMB temperature anisotropies and weak lensing shear. In order to speed up computations, it can be simplified by resorting to the Limber approximation \cite{Limber:1953, Kaiser:1992, Kaiser:1998} which gives
\begin{multline}
C_{\ell}^{\gamma\mathrm{T}} = \sqrt{\frac{(\ell+2)!}{(\ell-2)!}} 2 \,T_{\mathrm{CMB}} \int_{0}^{z_{*}} \mathrm{d}z \frac{\mathrm{d}}{\mathrm{d}z}[D(z)(1+z)] \\
 \times D(z)(1+z) \frac{g\boldsymbol{\left(}\chi(z)\boldsymbol{\right)}}{\chi^{2}(z)} P^{\mathrm{lin}}_{\Phi \Phi}\left(k=\frac{\ell+\sfrac{1}{2}}{\chi(z)}, z=0\right).
\label{eq:clsiswshear3}
\end{multline}
The power spectrum of the gravitational potential at late times is related to the matter power spectrum through Poisson's equation
\begin{equation}
P^{\mathrm{lin}}_{\Phi \Phi}(k, z=0) = \left(\frac{3}{2}\right)^{2} \frac{\Omega^{2}_{\mathrm{m}}H_{0}^{4}}{c^{4}} \frac{P^{\mathrm{lin}}_{\delta \delta}(k, z=0)}{k^{4}}.
\label{eq:poisson}
\end{equation}
For large $\ell$ we can make the approximations
\begin{equation}
\begin{aligned}
\sqrt{\frac{(\ell+2)!}{(\ell-2)!}} \sim \ell^{2}, \\
(\ell + \sfrac{1}{2})^{2} \sim \ell^{2}.
\label{eq:ellapprox}
\end{aligned}
\end{equation}
Using Equations \ref{eq:poisson} and \ref{eq:ellapprox} we can write Eq.~\ref{eq:clsiswshear3} as
\begin{widetext}
\begin{equation}
C^{\gamma\mathrm{T}}_{\ell} = 3 \frac{\Omega_{\mathrm{m}} H^{2}_{0}T_{\mathrm{CMB}}}{c^{2}} \frac{1}{(\ell+\sfrac{1}{2})^{2}} \int \mathrm{d}z \frac{\mathrm{d}}{\mathrm{d}z} \left[D(z)(1+z)\right]
D(z) W^{\gamma}\boldsymbol{\left(}\chi(z)\boldsymbol{\right)} P^{\mathrm{lin}}_{\delta \delta}\left(k=\frac{\ell+\sfrac{1}{2}}{\chi(z)}, 0\right),
\end{equation}
\end{widetext}
which is the expression given in Eq.~\ref{eq:clisw}.

\section{\label{sec:galaxysys} Treatment of systematic uncertainties in galaxy clustering data}

The number density of galaxies observed in SDSS DR8 photometric data is affected by various systematic uncertainties such as stellar density, Galactic extinction and PSF size variation \cite{Ross:2011, Ho:2012}. These effects remain even after masking and removal of the highest contamination regions. In order to obtain an unbiased galaxy overdensity map, we need to correct for the number density variation due to systematics. The SDSS recorded the values of several potential systematic uncertainties for the observed fields: airmass, Galactic extinction and seeing (as measured by the PSF FWHM) in all 5 SDSS bands for the field each galaxy has been observed in as well as sky emission at the position of the galaxy for all the 5 bands. These quantities can be queried for each galaxy position on the CAS\footnote{See footnote \ref{footn:cas}.}. In this work, we consider four different observational systematics: Galactic extinction in the $\tt{r\text{-}band}$ as well as FWHM of the PSF, airmass and sky emission in the $\tt{i\text{-}band}$. A further potential systematic uncertainty is the presence of foreground stars. \citet{Ross:2011} show that the effects of foreground stars on galaxy number density are largely independent of the magnitude of the stars. We therefore follow \citet{Ho:2012} and investigate how the number density of stars with $\tt{i\text{-}band}$ magnitudes in the range $18.0 \leq i < 18.5$ affects the number density of detected galaxies. 

\begin{figure*}
\begin{center}
\includegraphics[scale=0.4]{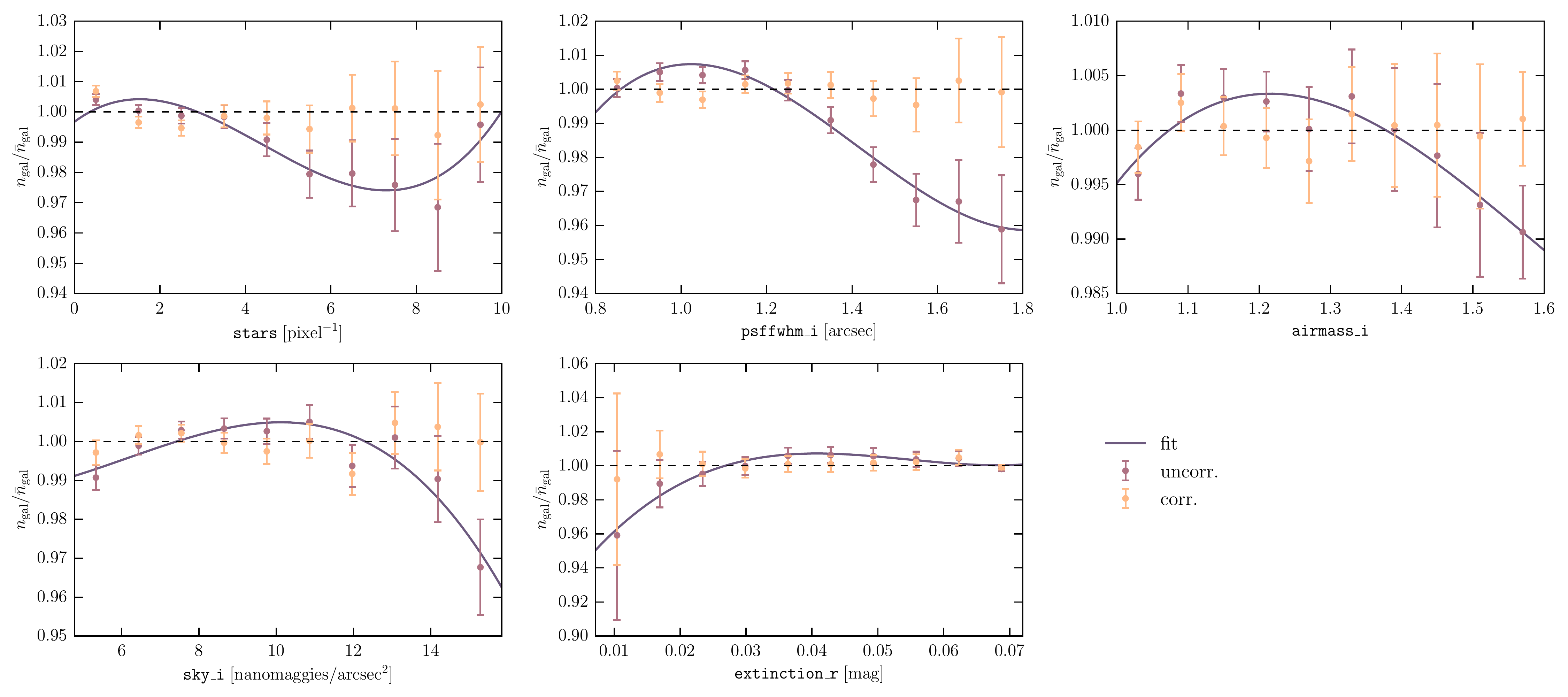}
\caption{Galaxy number density $n_{\mathrm{gal}}$ relative to mean galaxy density $\bar{n}_{\mathrm{gal}}$ as a function of potential systematic value. The figures show both the uncorrected data and the data corrected using a third order polynomial fit to the uncorrected relation. The error bars assume Poisson noise and are thus likely underestimated due to the correlations between galaxy positions.} 
\label{fig:galaxysys}
\end{center}
\end{figure*}

We pixelise all quantities onto HEALPix maps of resolution $\tt NSIDE$ $= 1024$ and compute the number density of galaxies relative to their mean number density as a function of the value of the systematic in the pixel. In order to correct for these systematic uncertainties, we fit a $3^{\mathrm{rd}}$-order polynomial to the functional dependence of the relative galaxy number density on the systematic. Then we multiply the uncorrected number densities by the inverse of this function. Various potential systematics such as Galactic extinction and stellar density are spatially correlated to one another. When correcting for various systematics simultaneously, the order in which the corrections are applied could influence results \cite{Ross:2011}. In our sample we find that the corrections are both independent of ordering and SDSS band and correcting for the effect in one band simultaneously corrects for all the other bands. The results are shown in Fig.~\ref{fig:galaxysys} and we use those to correct the galaxy maps from residual systematic uncertainties. We clip the systematics maps at the minimum and maximum systematics value shown in the figure and apply the fitted corrections to the galaxy number density. The galaxy clustering spherical harmonic power spectra before and after correcting for systematic uncertainties are discussed in Section \ref{subsec:deltagcls} and shown in Appendix \ref{sec:cltests}.

\section{\label{sec:cosmicshearanalysis}PSF correction and construction of weak lensing shear maps}

\begin{figure*}
\begin{center}
\includegraphics[scale=0.6]{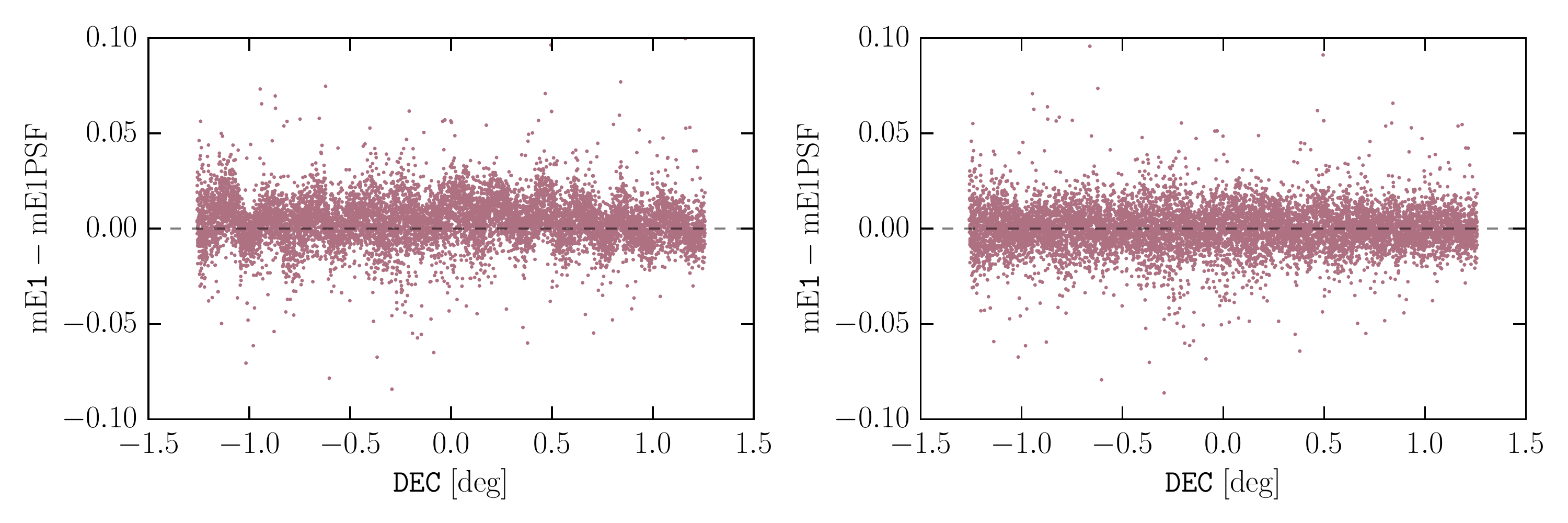}
\caption{Residuals between ellipticity component $e_{1}$ of a random sample of $10\,000$ bright stars measured on the co-add images and PSF models for these objects as a function of declination ($\tt DEC$) before (left panel) and after (right panel) applying the correction described in the text.} 
\label{fig:PSFmodcorr}
\end{center}
\end{figure*}

The galaxy shapes measured from images represent a convolution of the intrinsic galaxy shapes with the PSF of the telescope and the atmosphere. We therefore need to correct this effect using PSF estimates measured from the shapes of stars observed in the survey. As described in \citet{Annis:2014} and \citet{Lin:2012}, the PSF model for SDSS Stripe 82 data is derived from weighted sums of shapes measured in the individual runs as opposed to the co-adds. \citet{Lin:2012} found that this leads to biases that need to be removed prior to PSF correction. In order to correct for these effects, we follow the steps outlined in \citet{Lin:2012}. We select bright stars with $\tt{i\text{-}band}$ magnitudes in $16 < i < 17$, which pass the clean photometry cuts\footnote{See footnote \ref{footn:flags}.}, and fit polynomials to the residuals between their shapes measured from the co-adds and the PSF model for these stars. The residuals before and after subtraction of the polynomial fit are shown in Fig.~\ref{fig:PSFmodcorr}. We see that the correction introduced in \citet{Lin:2012} has considerably removed both an overall bias as well as discontinuities at the CCD camera column (camcol) edges. 

Using the revised PSF model, we correct the measured shapes for the effect of the PSF. We use the linear PSF correction algorithm derived in \citet{Hirata:2003}, which can be applied to the adaptive moment measurements from the SDSS $\tt PHOTO$ pipeline\footnote{Note that there is a typo in \citet{Hirata:2003}: The quantities $C_{g}, C_{f}, D_{g}, D_{f}$ in Eq. (B9) should be squared.}. 

In order to obtain a galaxy sample for reliable weak lensing shear measurement, we follow \citet{Lin:2012} and perform two additional selection cuts on the galaxies after PSF correction: we select galaxies with ellipticity components $e_{1}, e_{2}$ satisfying $\vert e_{1} \vert < 1.4$ as well as $\vert e_{2} \vert < 1.4$ and photometric redshift uncertainties $\sigma_{z} < 0.15$. This additional selection leaves a galaxy sample consisting of $N_{\mathrm{gal}} =  3\,322\,915$ galaxies.

\citet{Lin:2012} found a camcol dependent additive bias in the PSF-corrected ellipticities. The mean ellipticities for each camcol lie in the range $\vert \bar{e}_{1} \vert = [6\times10^{-5}, 0.02]$ and $\vert \bar{e}_{2} \vert = [0.002, 0.009]$, which is larger than expected for a mean zero field \cite{Lin:2012}. We therefore follow \citet{Lin:2012} and correct for the additive bias by subtracting the mean ellipticity for each camcol. We choose to perform this step prior to coordinate transformation (i.e. for ellipticities defined relative to equatorial coordinates) as opposed to after rotation. We find that removing the mean camcol ellipticity reduces PSF leakage to a level which is subdominant in our analysis.

\begin{figure*}
\begin{center}
\includegraphics[scale=0.6]{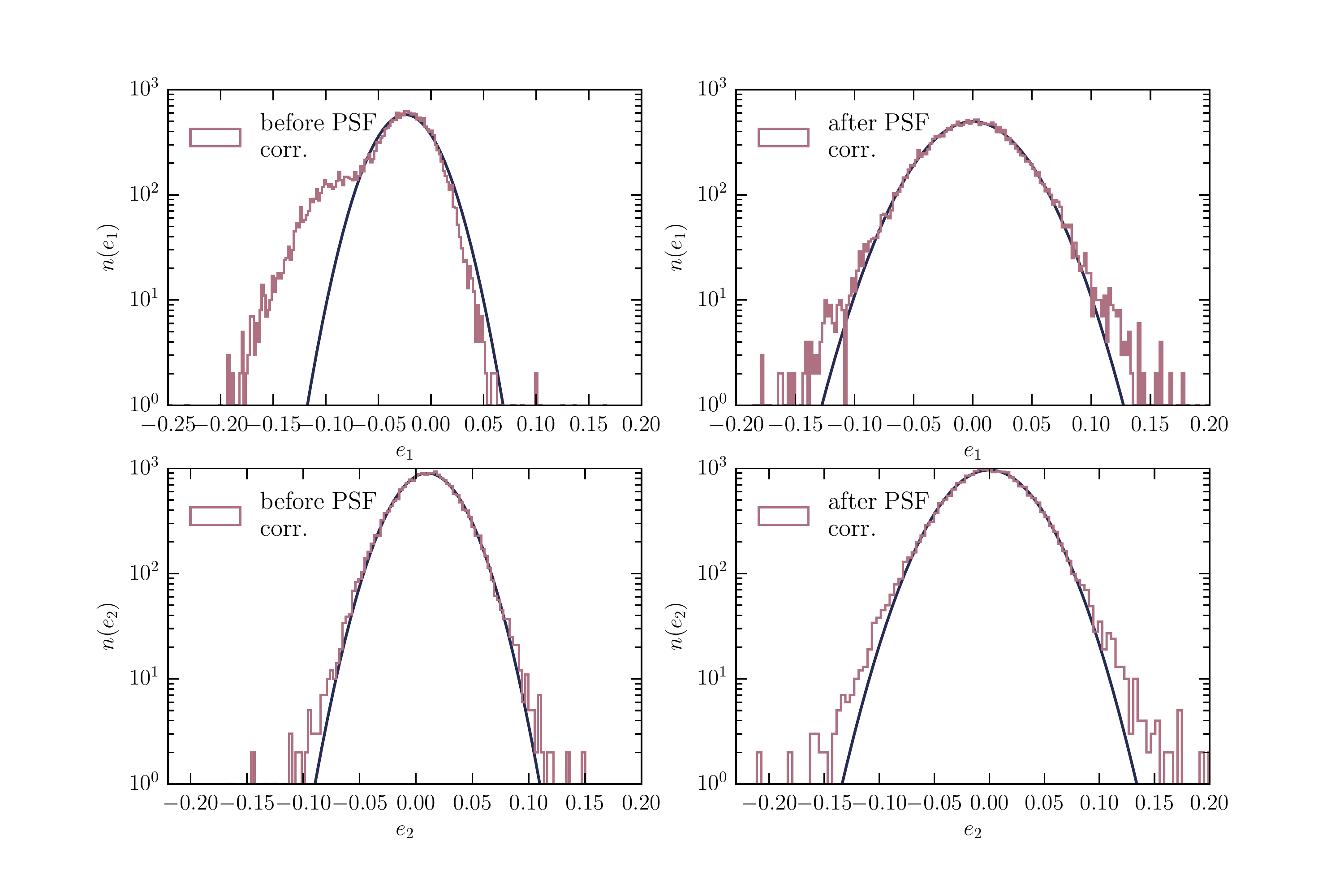}
\caption{Histograms of the ellipticity components $e_{1}$ and $e_{2}$ averaged over HEALPix pixels of resolution $\tt NSIDE$ $= 512$. The left panels show the distributions before correction for the PSF and subtraction of a camcol dependent additive bias, while the right panels show the distributions after the application of these corrections.} 
\label{fig:elliphist} 
\end{center}
\end{figure*}

Fig.~\ref{fig:elliphist} shows the distributions of the ellipticity components $e_{1}$ and $e_{2}$ defined relative to equatorial coordinates. They are averaged over HEALPix pixels of resolution $\tt NSIDE$ $= 512$, which corresponds to a pixel area of $A_{\mathrm{pix}} \approx 0.013$ deg$^{2}$. The figure displays the ellipticity histograms both prior to PSF correction and subtraction of additive bias as well as the final distributions obtained after applying both corrections. We see that the corrections have removed the effects of the PSF and the final histograms can be described by Gaussian distributions.

In the final step, these ellipticities need to be transformed to shear estimates by correcting for the shear resolution factor $\mathcal{R}$, which is defined as
\begin{equation}
\mathcal{R} = \left \langle \frac{\partial \hat{\gamma}_{i}}{\partial \gamma_{i}}\right \rangle.
\end{equation}
The shear resolution factor $\mathcal{R}$ quantifies the response of the estimated mean ellipticity to an applied shear. For the adaptive moment method described in \citet{Bernstein:2002} it is given by $\mathcal{R} = 2(1 - e^{2}_{\mathrm{int}})$, where $e_{\mathrm{int}}$ denotes the intrinsic r.m.s. ellipticity per component. We follow \citet{Lin:2012} and use $e_{\mathrm{int}} = 0.37$ as measured by \citet{Hirata:2004}. 

In order to construct the final weak lensing shear maps we thus apply the resolution correction to the ellipticities and transform them from equatorial to Galactic coordinates.
  
\section{\label{sec:shearrotation} Transformation of weak lensing shear under rotation}

We rotate the weak lensing galaxy shears from equatorial to Galactic coordinates following the implementation in HEALPix. The method is briefly summarised below.

The rotation angle of the shears under a coordinate rotation as described by the rotation matrix $R$ is equal to twice the rotation angle $\psi$ of the coordinate axes with respect to which they are defined. In HEALPix the x-axis is in the direction of $\mathbf{e}_{\phi}$ and the y-axis in the direction of $\mathbf{e}_{\theta}$. 

In order to derive $\psi$ we define the following quantities: The position before rotation is denoted as $\mathbf{r} = \begin{pmatrix} x & y & z \end{pmatrix}$ and the position after rotation is $\mathbf{r}' = \begin{pmatrix} x' & y' & z' \end{pmatrix}$. We further define the vector towards the north pole in the unrotated coordinate system, which is given by $\mathbf{p} = \begin{pmatrix} x_{0} & y_{0} & z_{0} \end{pmatrix} = \begin{pmatrix} 0 & 0 & 1 \end{pmatrix}$. Under the inverse rotation $R^{-1}$ the north polar vector is mapped to $\mathbf{p}'' = \begin{pmatrix} x''_{0} & y''_{0} & z''_{0} \end{pmatrix}$. At the position $\mathbf{r}$ the unit vectors in $\theta$- and $\phi$-direction are given by
\begin{equation}
\begin{aligned}
\mathbf{e}_{\phi} &= \frac{\mathbf{p} \times \mathbf{r}}{\vert \mathbf{p} \times \mathbf{r} \vert}, \\
\mathbf{e}_{\theta} &= \frac{(\mathbf{p} \times \mathbf{r}) \times \mathbf{r}}{\vert (\mathbf{p} \times \mathbf{r}) \times \mathbf{r} \vert}.
\end{aligned}
\end{equation}
We have the following identities
\begin{equation}
\begin{aligned}
R \, \mathbf{e}_{\phi \backslash \theta} \cdot \mathbf{p} &= \mathbf{e}_{\phi \backslash \theta} \cdot R^{-1} \mathbf{p}, \\
R^{-1} \mathbf{p} \cdot \mathbf{r} &= \mathbf{p} \cdot R \, \mathbf{r}.
\end{aligned}
\end{equation}
Taking into account the left-handedness of the HEALPix coordinate system and inserting the explicit expressions, it follows that 
\begin{equation}
\begin{aligned}
\cos{\psi} &= \frac{c}{\sqrt{1-z^{2}}}(z z' - z''_{0}), \\
\sin{\psi} &= \frac{c}{\sqrt{1-z^{2}}}(x y''_{0} - y x''_{0}),
\end{aligned}
\end{equation}
where $c$ is a constant, which we can remove by ensuring that $\sin^{2}{\psi}+\cos^{2}{\psi} = 1$. Under this rotation the weak lensing shear transforms as
\begin{equation}
\begin{aligned}
\gamma'_{1} &= \cos{2\psi} \, \gamma_{1} + \sin{2\psi} \, \gamma_{2}, \\
\gamma'_{2} &= -\sin{2\psi} \, \gamma_{1} + \cos{2\psi} \, \gamma_{2}.
\end{aligned}
\end{equation}

\section{\label{sec:maskcorr} Choice of $\tt{PolSpice}$ parameter settings}

\begin{figure}
\includegraphics[scale=0.6]{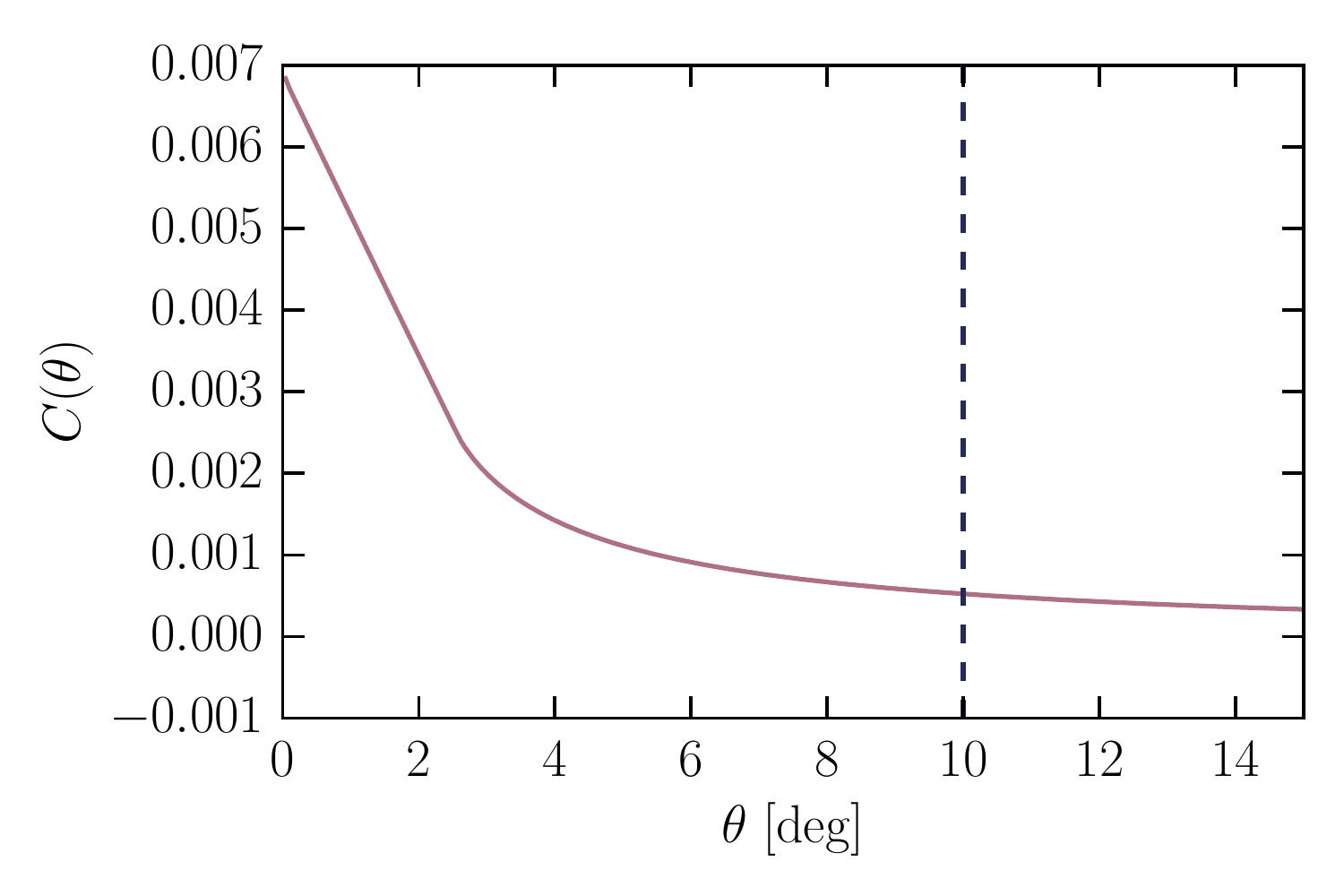}
\caption{Real space correlation function of the SDSS Stripe 82 mask. The dashed line denotes the value chosen for $\theta_{\mathrm{max}}$.}
\label{fig:maskcorrfunc}
\end{figure}

In this section we illustrate the determination of the maximal angular scale $\theta_{\mathrm{max}}$ used to compute spherical harmonic power spectra on the example of the SDSS Stripe 82 mask. Fig.~\ref{fig:maskcorrfunc} shows the real space correlation function of this mask. It is non-zero for small angular scales, then starts to fall-off and approximately vanishes for large angular separations. From this figure we see that $\theta_{\mathrm{max}}$ is not a well-defined quantity. Our approach is thus to choose a maximal angular scale by eye and validate it on the Gaussian simulations. For the SDSS Stripe 82 mask, we choose $\theta_{\mathrm{max}} = 10$ degrees and $\theta_{\mathrm{FWHM}} = 5$ degrees. When testing these $\tt{PolSpice}$ settings on the simulations, we find a reasonable agreement between input and recovered power spectra. 

\section{\label{sec:corrmaps}Correlated maps of spin-0 and spin-2 fields}

Our analysis relies on Gaussian simulations both for validation of the data analysis pipeline and covariance matrix estimation. We thus need to generate correlated HEALPix maps of both spin-0 and spin-2 fields from input auto- and cross-power spectra. \citet{Cabre:2007} and \citet{Giannantonio:2008} describe an algorithm for generating correlated HEALPix maps of spin-0 fields. In order to consistently simulate the weak lensing shear field, we extend this algorithm to also include correlations between spin-0 and spin-2 fields. 

These algorithms are all based on the HEALPix routine $\tt synfast$, which generates HEALPix maps of realisations of input spherical harmonic power spectra $C_{\ell}^{ii}$. If the fields are additionally mean-subtracted, this is equivalent to requiring that the spherical harmonic coefficients $a_{\ell m}$ of the maps satisfy
\begin{equation}
\begin{aligned}
\left \langle a^{i}_{\ell m} \right \rangle &= 0, \\
\left \langle a^{i}_{\ell m} a^{i *}_{\ell' m'} \right \rangle &= C_{\ell}^{ii} \, \delta_{\ell \ell'} \, \delta_{m m'}.
\end{aligned}
\end{equation} 
In $\tt synfast$ these conditions are imposed by assigning a random phase $\xi$ with mean 0, $\langle \xi \rangle = 0$, and unit variance, $\langle \xi \xi^{*} \rangle = 1$, to each spherical harmonic mode $\ell$ and setting
\begin{equation}
a^{i}_{\ell m} = \sqrt{C_{\ell}^{ii}} \xi.
\end{equation} 
As derived in \citet{Giannantonio:2008}, this method can be extended to correlated maps using more random phases. The simplest case is to create two correlated spin-0 zero maps with power spectra $C_{\ell}^{ii}$, $C_{\ell}^{jj}$ and cross-power spectrum $C_{\ell}^{ij}$. This is the only case relevant for our work and it is achieved by choosing the amplitudes of the maps of the two probes $i, \,j$ as \cite{Giannantonio:2008}
\begin{equation}
\begin{aligned}
a^{i}_{\ell m} &= \sqrt{C_{\ell}^{ii}} \xi_{1}, \\
a^{j}_{\ell m} &= \frac{C_{\ell}^{ij}}{\sqrt{C_{\ell}^{ii}}} \xi_{1} + \sqrt{C_{\ell}^{jj} - \frac{(C_{\ell}^{ij})^{2}}{C_{\ell}^{ii}}}\xi_{2}.
\end{aligned}
\label{eq:corrgauss}
\end{equation}
As described in \cite{Giannantonio:2008} this algorithm can be implemented using $\tt synfast$ by first creating a map with power spectrum $C_{\ell}^{ii}$ and a second map using the same seed with power spectrum $\sfrac{(C_{\ell}^{ij})^{2}}{C_{\ell}^{ii}}$. Finally the second map needs to be added to a third map, generated with a different random seed and with power spectrum $C_{\ell}^{jj} - \sfrac{(C_{\ell}^{ij})^{2}}{C_{\ell}^{ii}}$. This ensures the desired auto- and cross-correlations. 

To extend this algorithm to spin-2 fields, we make use of the polarisation version of $\tt synfast$, which allows us to generate correlated spin-0 and spin-2 maps consistent with input auto- and cross-power spectra. Let $0$ denote the spin-0 field. Then $C^{00}_{\ell}$ denotes the auto-power spectrum of the spin-0 field, $C^{\mathrm{EE}}_{\ell}$, $C^{\mathrm{BB}}_{\ell}$ are the E- and B-mode power spectra of the spin-2 field and $C^{0\mathrm{E}}_{\ell}$ is the cross-power spectrum between the spin-0 field and the spin-2 E mode. Given these input power spectra, the polarisation mode of $\tt synfast$ generates a map of the spin-0 field and two maps of the spin-2 field with the desired auto- and cross-power spectra. 

In order to obtain correlated maps $m_{\mathrm{T}}, m_{\delta_{g}}, m_{\gamma_{1}}, m_{\gamma_{2}}$ of CMB temperature anisotropies, galaxy overdensity and galaxy weak lensing shear with auto- and cross-power spectra $C^{\mathrm{TT}}_{\ell}, C_{\ell}^{\delta_{g}\mathrm{T}}, C_{\ell}^{\delta_{g} \delta_{g}}, C^{\gamma\mathrm{T}}_{\ell}, C^{\gamma\delta_{g}}_{\ell}, C^{\gamma \gamma}_{\ell}$ we therefore proceed as follows:
\begin{enumerate}[label=(\roman*)]
\item We first create three correlated HEALPix maps using $\tt synfast$ in polarisation mode with the power spectra 
\begin{align*}
C^{00}_{\ell} &= C^{\mathrm{TT}}_{\ell},\\
C^{\mathrm{EE}}_{\ell} &= \sfrac{C^{\gamma \gamma}_{\ell}}{2},\\
C^{\mathrm{BB}}_{\ell} &= 0,\\
C^{0\mathrm{E}}_{\ell} &= C^{\gamma\mathrm{T}}_{\ell}.\\
\end{align*}
These maps are denoted $m^{1}_{i}$, where $i \in \{\mathrm{T}, \gamma_{1}, \gamma_{2}\}$.
\item Following Eq.~\ref{eq:corrgauss}, we then create three maps with a new random seed and the power spectra
\begin{align*}
C^{00}_{\ell} &= C_{\ell}^{\delta_{g} \delta_{g}} - \sfrac{(C_{\ell}^{\delta_{g}\mathrm{T}})^{2}}{C_{\ell}^{\mathrm{TT}}},\\
C^{\mathrm{EE}}_{\ell} &= \sfrac{C^{\gamma \gamma}_{\ell}}{2},\\
C^{\mathrm{BB}}_{\ell} &= 0,\\
C^{0\mathrm{E}}_{\ell} &= C^{\gamma\delta_{g}}_{\ell}.\\
\end{align*}
These maps are denoted $m^{2}_{i}$, where $i \in \{\delta_{g}, \gamma_{1}, \gamma_{2}\}$.
\item We create a spin-0 map generated with the same seed as used for $m^{1}$ with the power spectrum
\begin{equation*}
C^{00}_{\ell} = \sfrac{(C_{\ell}^{\delta_{g}\mathrm{T}})^{2}}{C_{\ell}^{\mathrm{TT}}},
\end{equation*}
which is called $m^{3}$.
\item Finally we combine the maps i.e.
\begin{align*}
m_{\mathrm{T}} &= m^{1}_{\mathrm{T}}, \\
m_{\delta_{g}} &= m^{2}_{\delta_{g}} + m^{3}, \\
m_{\gamma_{1}} &= m^{1}_{\gamma_{1}} + m^{2}_{\gamma_{1}}, \\
m_{\gamma_{2}} &= m^{1}_{\gamma_{2}} + m^{2}_{\gamma_{2}}.
\end{align*}
\end{enumerate}
This procedure yields four correlated maps with auto- and cross-power spectra $C^{\mathrm{TT}}_{\ell}, C_{\ell}^{\delta_{g}\mathrm{T}}, C_{\ell}^{\delta_{g} \delta_{g}}, C^{\gamma\mathrm{T}}_{\ell}, C^{\gamma\delta_{g}}_{\ell}, C^{\gamma \gamma}_{\ell}$. The algorithm described above introduces an unwanted, additional correlation between $m_{\delta_{g}}$ and $m_{\gamma_{1}}, m_{\gamma_{2}}$. It can in principle be corrected for by adding counter-terms to the respective maps. Since the additional correlation is subdominant in the present case, we neglect these counter-terms.

In order to obtain realistic maps we need to account for the effects of HEALPix pixel and beam window function. The signal measured in each HEALPix pixel is a convolution of the underlying signal with the HEALPix window function. If further experimental beams are present, the signal is additionally convolved with the beam window function. Since a convolution in real space is equivalent to a multiplication in Fourier space, we account for these effects by multiplying the input power spectra by the power spectra of the respective window functions prior to generating the HEALPix maps. 

To compute the covariance matrix as well as to validate the analysis pipeline we need to add realistic noise to the correlated Gaussian simulations. We choose to add the noise on the map level. For the CMB temperature anisotropies we add the $\tt Commander$ HMHD map provided by the Planck collaboration to each simulated temperature map. We do not randomise the noise map for each new realisation since the HMHD map features significant correlations which would be lost by randomising. Since we are adding the same noise map to each random realisation we expect to slightly underestimate the noise using our simulations. However, we do not expect this to have a significant effect on our results, since the noise in the CMB temperature power spectrum is dominated by cosmic variance at the scales considered. For the galaxy overdensity field we create noise maps by randomising the positions of the galaxies in our data inside the survey mask. We then pixelise those on a HEALPix map and add the noise map to the simulated map. The galaxy shear noise maps are created by rotating each galaxy shear by a random angle and repixelising the rotated shears onto HEALPix maps. As before these noise maps are added to the signal maps to produce the Gaussian simulations including both signal and noise. 

\section{\label{sec:validation}Validation of spherical harmonic power spectrum measurements}

We validate the spherical harmonic power spectrum measurement outlined in Section \ref{sec:cls} using the correlated Gaussian simulations described in Appendix \ref{sec:corrmaps}. We compute theoretical predictions for the six spherical harmonic power spectra considered in this work, i.e. $C^{\mathrm{TT}}_{\ell},\, C_{\ell}^{\delta_{g}\mathrm{T}}, \,C_{\ell}^{\delta_{g} \delta_{g}}, \,C^{\gamma\mathrm{T}}_{\ell}, \,C^{\gamma\delta_{g}}_{\ell}, \,C^{\gamma \gamma}_{\ell}$ for a $\Lambda$CDM cosmological model with parameters $\{h,\, \Omega_{\mathrm{m}}, \,\Omega_{\mathrm{b}}, \,\Omega_{\Lambda}, \,n_{\mathrm{s}}, \,\sigma_{8}, \,\tau_{\mathrm{reion}}, \,T_{\mathrm{CMB}}\} = \{0.7, \,0.3, \,0.049, \,1.0, \,0.88, \,0.078, \,2.275 \,\mathrm{K}\}$. We further set the linear, redshift-independent galaxy bias parameter to $b=2$.

Using the algorithm described above, we generate $N_{\mathrm{sim}}=1000$ Gaussian realisations of these power spectra and add the noise maps determined from the data. We then apply angular masks equivalent to those in the data and compute the spherical harmonic power spectra from the masked maps using the exact same methodology and $\tt{PolSpice}$ settings as applied on the data. To estimate the noise bias we follow the same randomisation approaches as outlined in Section \ref{sec:cls}.

\begin{figure*}
\begin{center}
\includegraphics[scale=0.4]{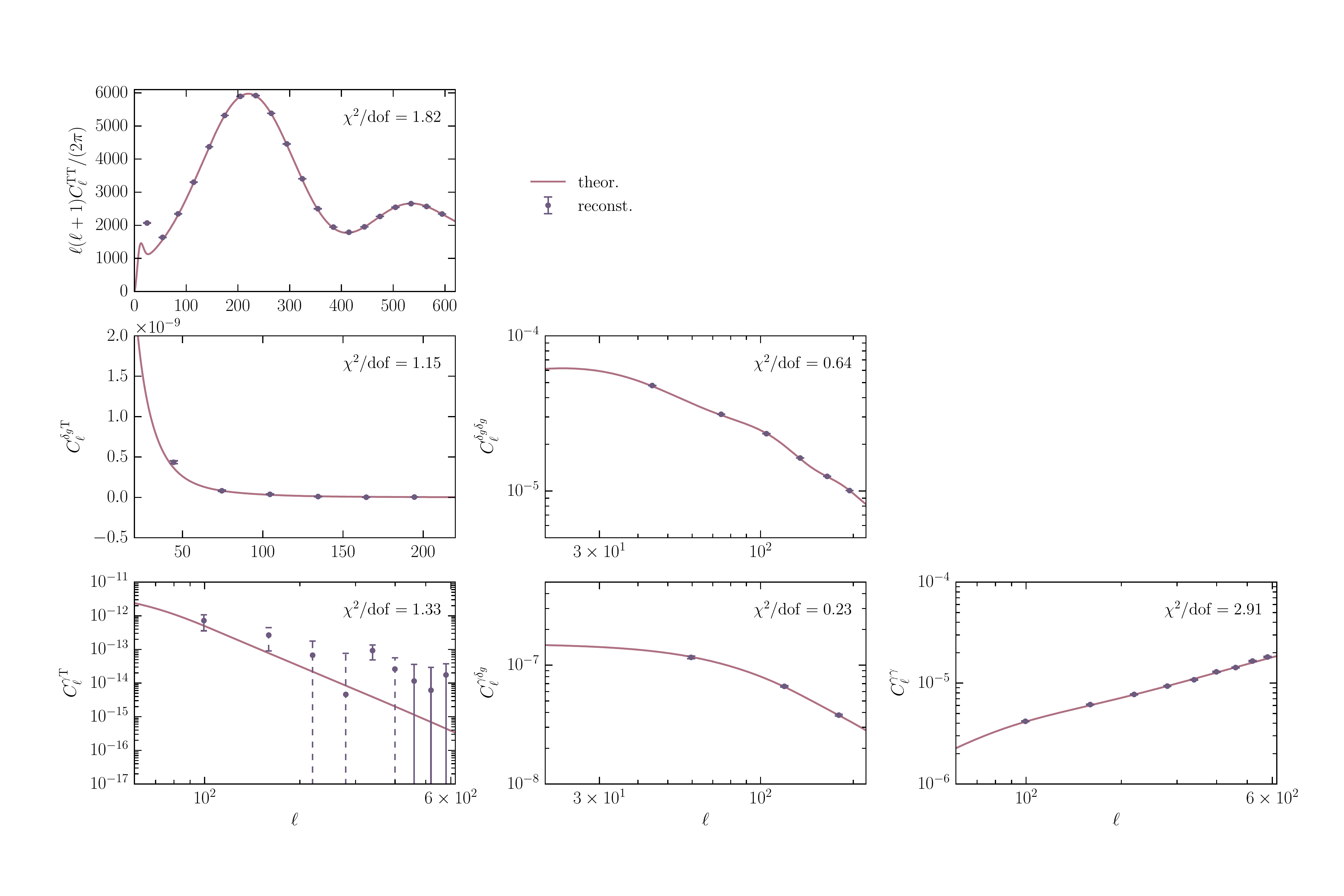}
\caption{Comparison between input power spectra and mean recovered power spectra as estimated from $N_{\mathrm{sim}}=1000$ Gaussian realisations generated using the algorithm outlined in Appendix \ref{sec:corrmaps}. The noise level of the Gaussian realisations is tuned to the data and the spherical harmonic power spectra have been computed using the same methodology and $\tt{PolSpice}$ settings as applied on the data. The angular multipole ranges and binning schemes for all power spectra are summarised in Table \ref{tab:clparams}. Dashed lines denote negative spherical harmonic power spectrum values.} 
\label{fig:mockrecon}
\end{center}
\end{figure*}

\begin{figure}
\includegraphics[scale=0.6]{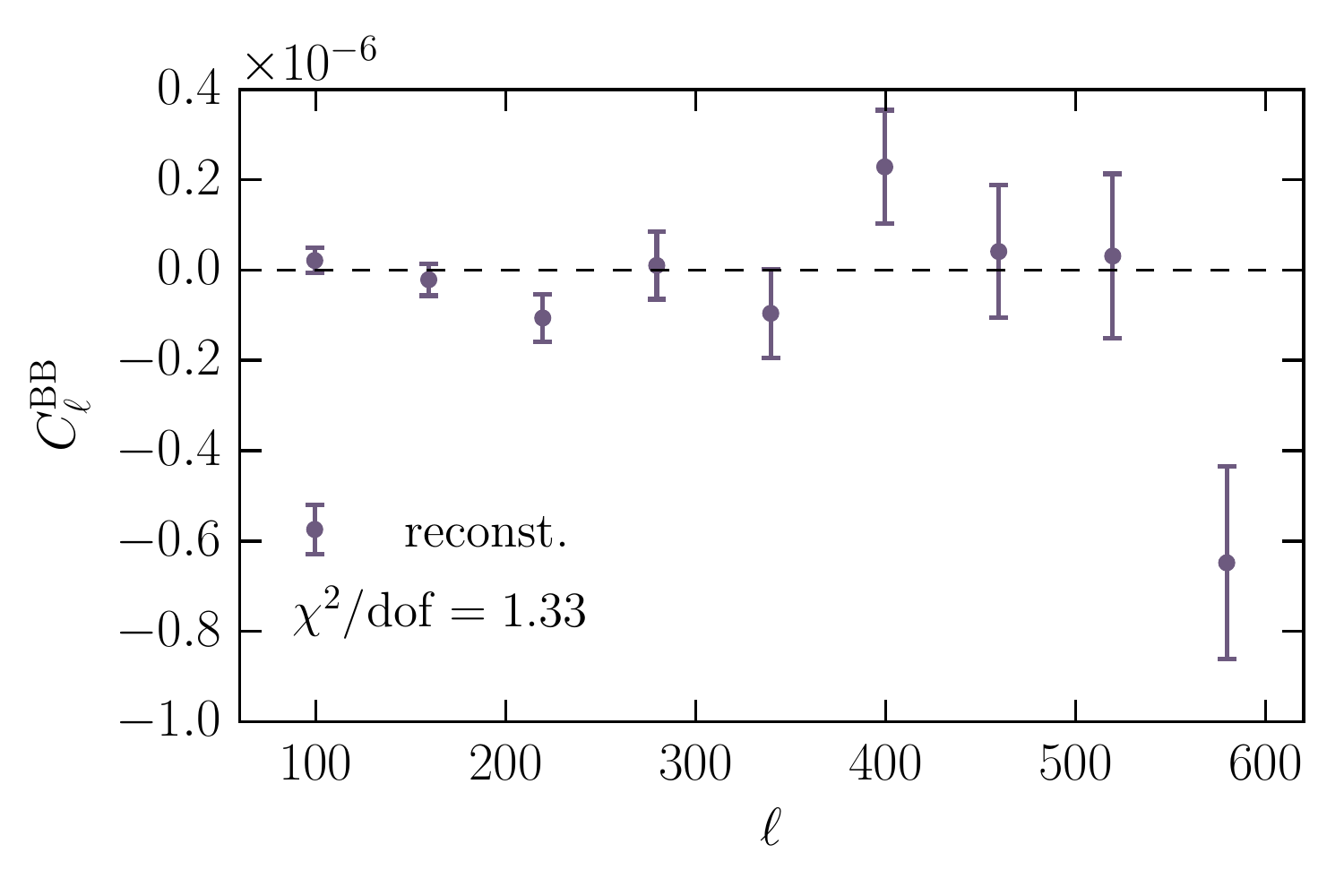}
\caption{The cosmic shear B-mode power spectrum reconstructed from $N_{\mathrm{sim}}=1000$ Gaussian realisations generated using the algorithm outlined in Appendix \ref{sec:corrmaps}. The angular multipole range and binning scheme is summarised in Table \ref{tab:clparams}.}
\label{fig:mockreconshearclsb}
\end{figure}

Figures \ref{fig:mockrecon} and \ref{fig:mockreconshearclsb} show a comparison between the input power spectra for all the six probes and the means of the recovered realisations. The uncertainties are derived from the sample covariance of the Gaussian realisations. We see that the input power spectra are recovered rather well. Also shown are the $\chi^{2}$ values between the reconstruction and the theory. These are not rigorous measures for the goodness of the recovery since they strongly depend on binning and angular multipole range chosen. 

\section{\label{sec:cltests}Spherical harmonic power spectrum robustness tests}

This section summarises the robustness tests performed for the spherical harmonic power spectra. 

\subsection{\label{subsec:eclgal} Comparison between spherical harmonic power spectra in equatorial and Galactic coordinates}

\begin{figure*}
\includegraphics[scale=0.6]{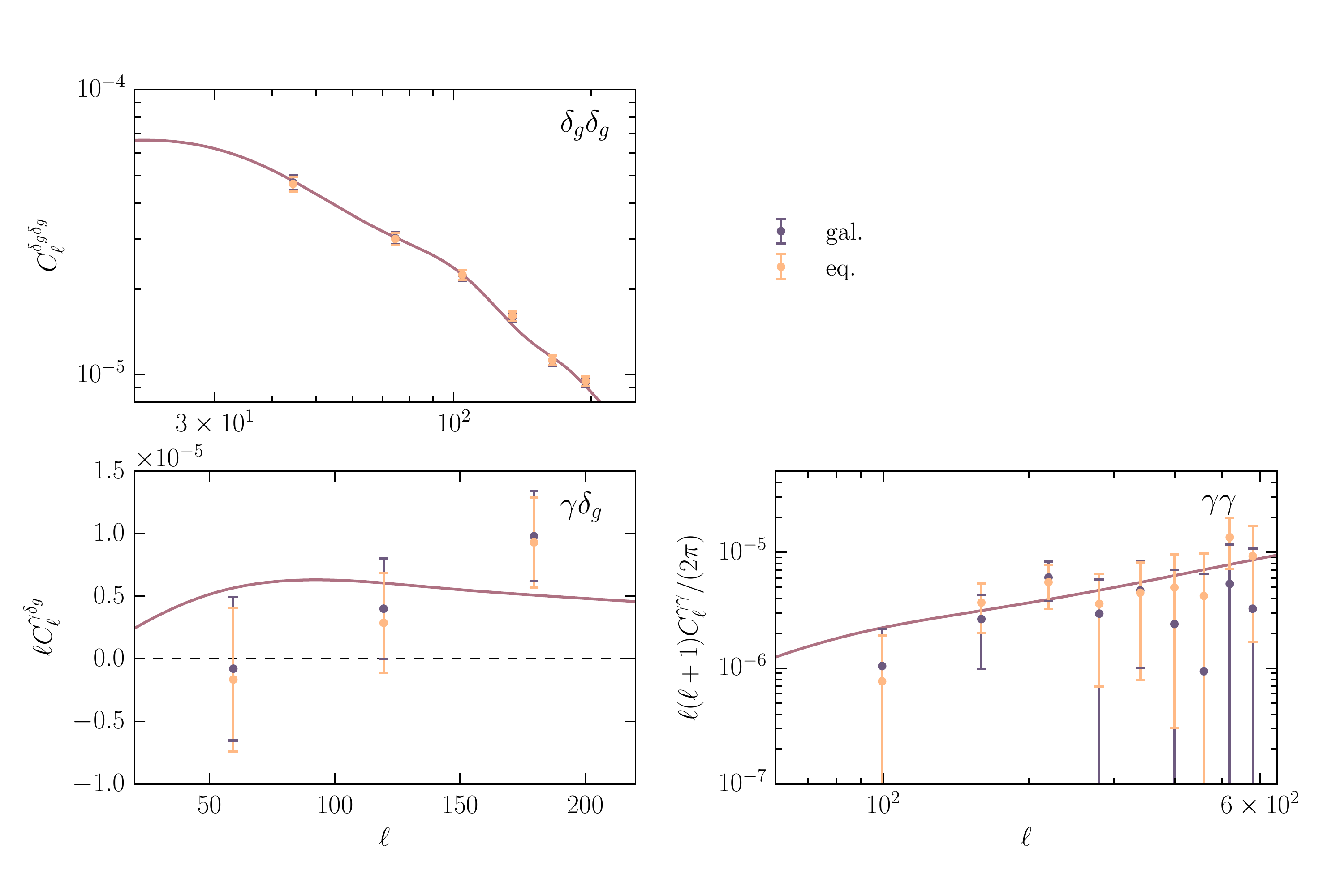}
\caption{Comparison between spherical harmonic power spectra computed from the maps in Galactic and equatorial coordinates.}
\label{fig:eclgalcls}
\end{figure*}

\begin{figure*}
\includegraphics[scale=0.6]{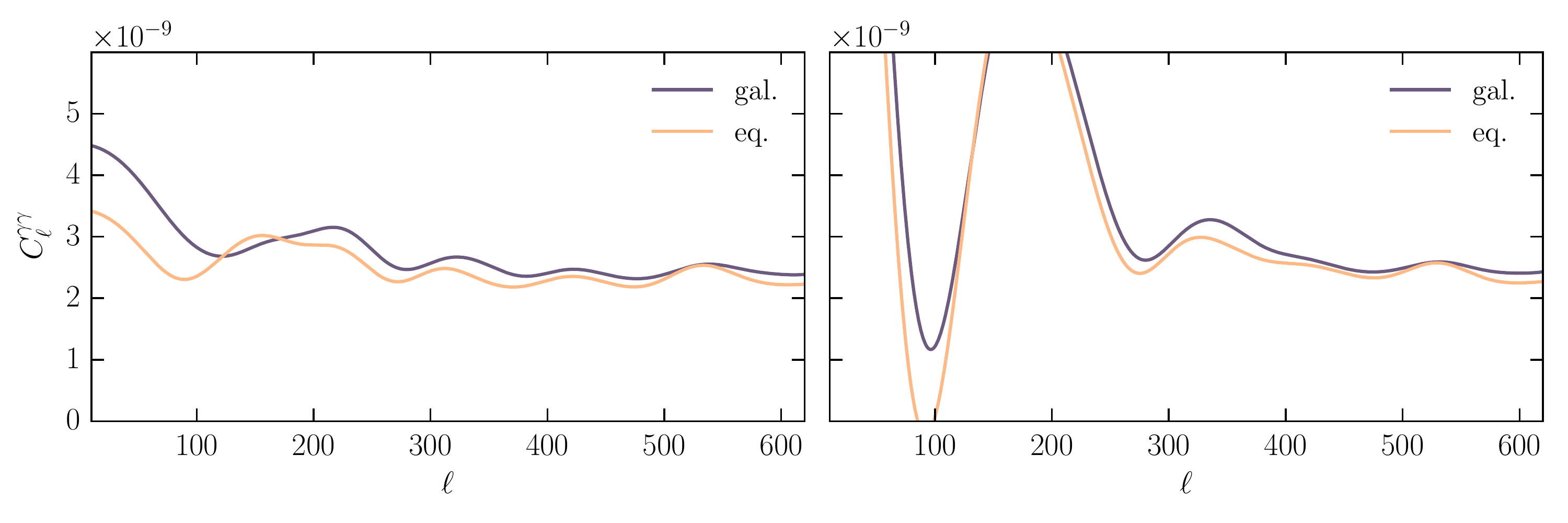}
\caption{Comparison between cosmic shear spherical harmonic power spectra prior to noise removal measured from the maps in Galactic and equatorial coordinates. The left hand panel shows the results when applying the correction for additive bias in equatorial coordinates and then rotating the shears to Galactic coordinates. The right hand panel shows the results when no additive bias correction is applied.}
\label{fig:compccnocc}
\end{figure*}

We test that the spherical harmonic power spectra involving maps which can be transformed between coordinate systems, i.e. galaxy overdensity and weak lensing shear maps, are unaffected by the rotation. The comparison between spherical harmonic power spectra computed from maps in Galactic and equatorial coordinates are shown in Fig.~\ref{fig:eclgalcls}. We find good agreement between the two power spectra for both $C^{\delta_{g}\delta_{g}}_{\ell}$ and $C^{\gamma\delta_{g}}_{\ell}$, while we find discrepancies for $C^{\gamma\gamma}_{\ell}$. We attribute this to the additive bias correction applied to the galaxy shears as outlined in Section \ref{subsec:gammacls}. The additive bias correction described in Appendix \ref{sec:cosmicshearanalysis}, causes an asymmetry between the galaxy shears in different coordinate systems, which is the cause for the large discrepancies detected. This can be seen from Fig.~\ref{fig:compccnocc}, which shows a comparison between the cosmic shear power spectra prior to noise removal as estimated from maps in Galactic and equatorial coordinates. The left panel shows the comparison when the additive bias correction is applied while in the right panel we do not apply any correction. As can be seen, we find discrepancies when we apply the additive bias correction in equatorial coordinates and then rotate the corrected shears to Galactic coordinates. Not applying any additive bias correction on the other hand, removes most of these effects. 

\subsection{\label{subsec:planckmaps} Comparison between spherical harmonic power spectra derived from different foreground-reduced CMB maps}

\begin{figure*}
\includegraphics[scale=0.6]{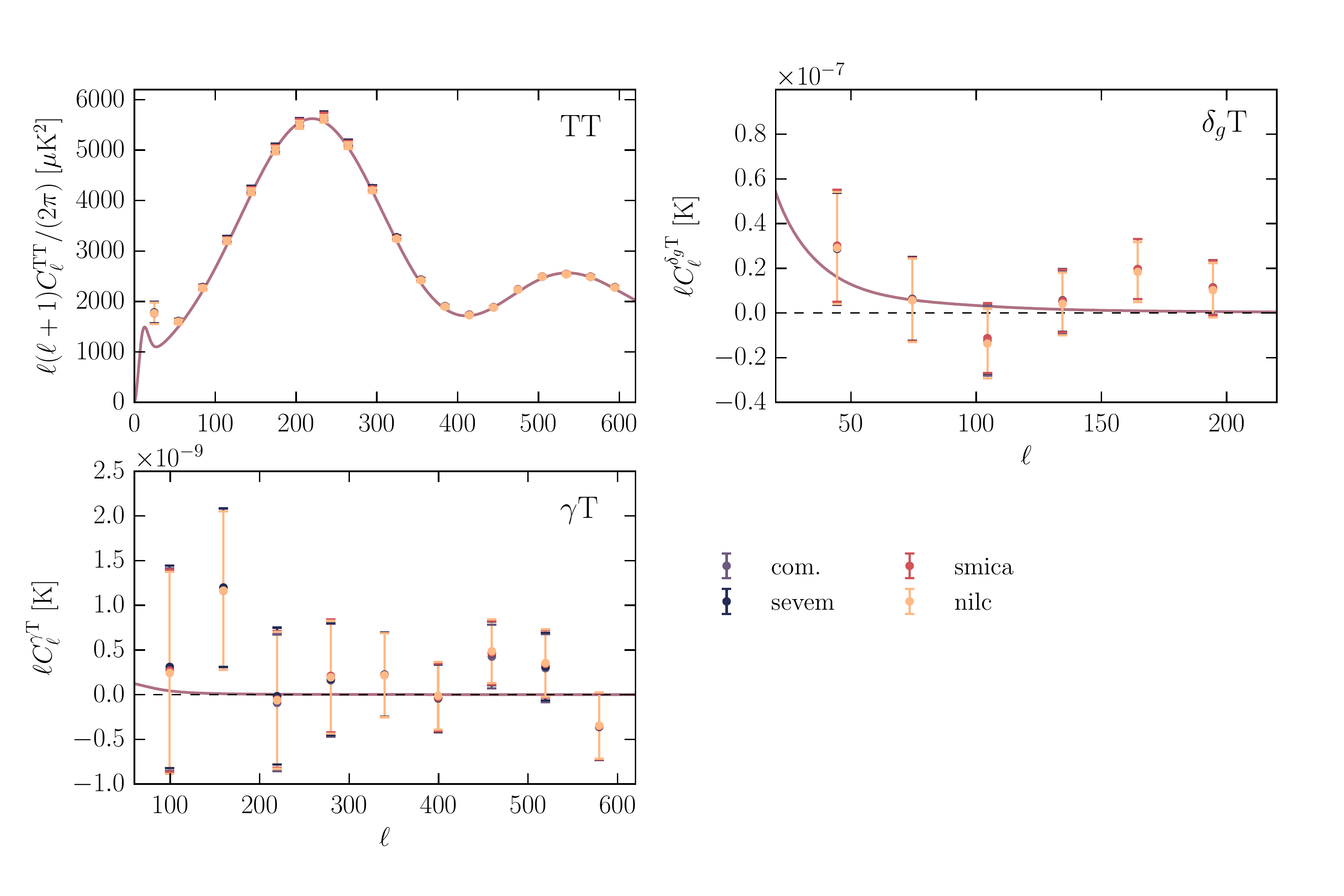}
\caption{Comparison between spherical harmonic power spectra $C^{\mathrm{TT}}_{\ell},\, C_{\ell}^{\delta_{g}\mathrm{T}}, \,C^{\gamma\mathrm{T}}_{\ell}$ derived using the four different foreground-reduced CMB maps from $\tt Commander$, $\tt NILC$, $\tt SEVEM$ and $\tt SMICA$.}
\label{fig:compplanckmaps}
\end{figure*}

We test that the spherical harmonic power spectra involving CMB data are unaffected by our choice of foreground-reduced map. The power spectra involving CMB data are shown in Fig.~\ref{fig:compplanckmaps} for the foreground-reduced CMB maps derived using the component separation methods $\tt Commander$, $\tt NILC$, $\tt SEVEM$ and $\tt SMICA$. As can be seen, the power spectra are virtually the same.

\subsection{\label{subsec:deltagforegroundcorr} Impact of systematics correction on galaxy clustering power spectrum}

\begin{figure}
\includegraphics[scale=0.6]{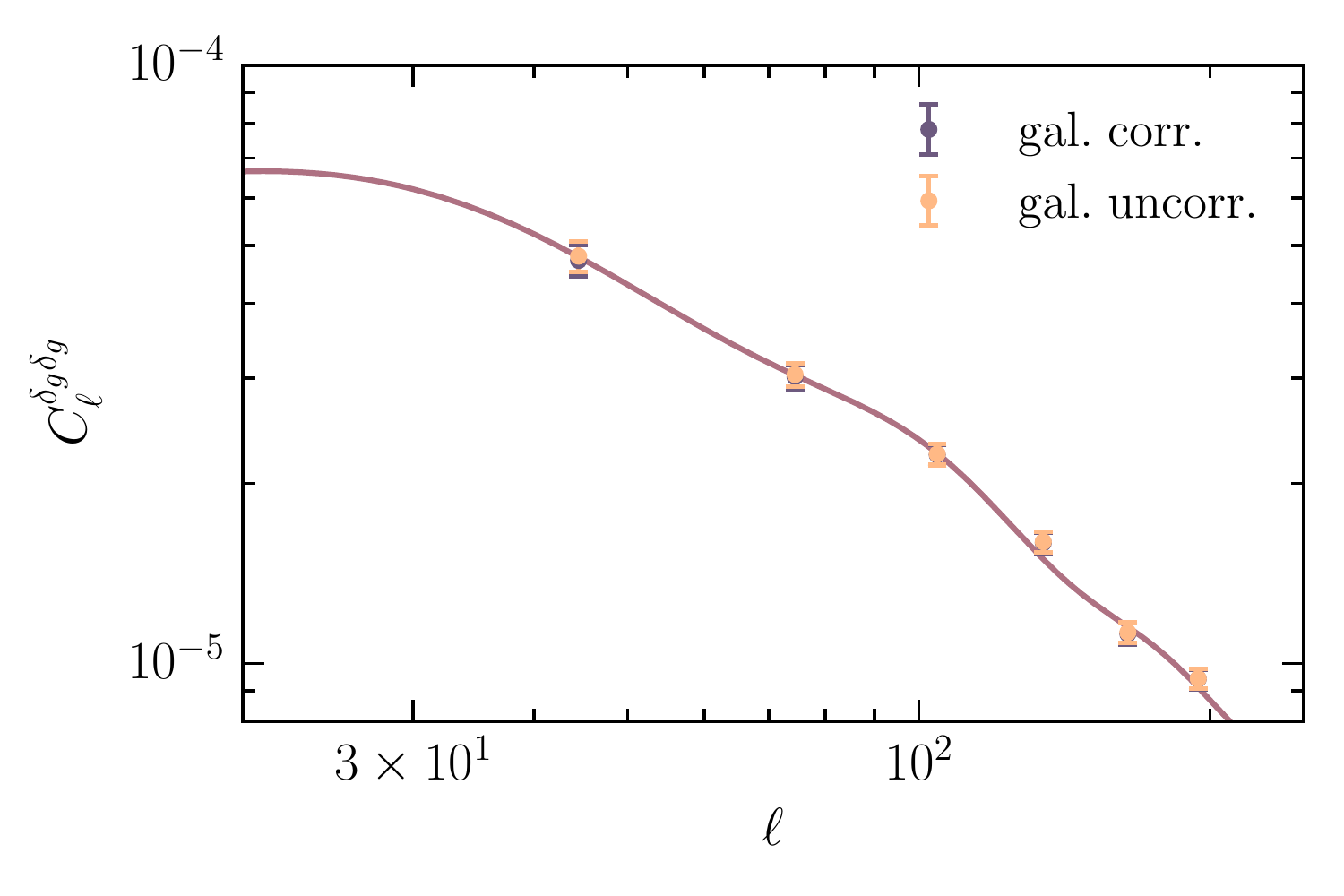}
\caption{Comparison between galaxy overdensity power spectra computed before and after systematics removal.}
\label{fig:deltagcorrnocorrcls}
\end{figure}

We further investigate the effect of systematics correction on the galaxy clustering power spectrum. The galaxy clustering spherical harmonic power spectra before and after correcting for systematic uncertainties are shown in Fig.~\ref{fig:deltagcorrnocorrcls}. Our systematics removal method slightly reduces the clustering amplitude at large angular scales, while leaving small angular scales almost unaffected. This is to be expected since Galactic foregrounds typically exhibit significant large scale clustering. 

\begin{figure}
\includegraphics[scale=0.5]{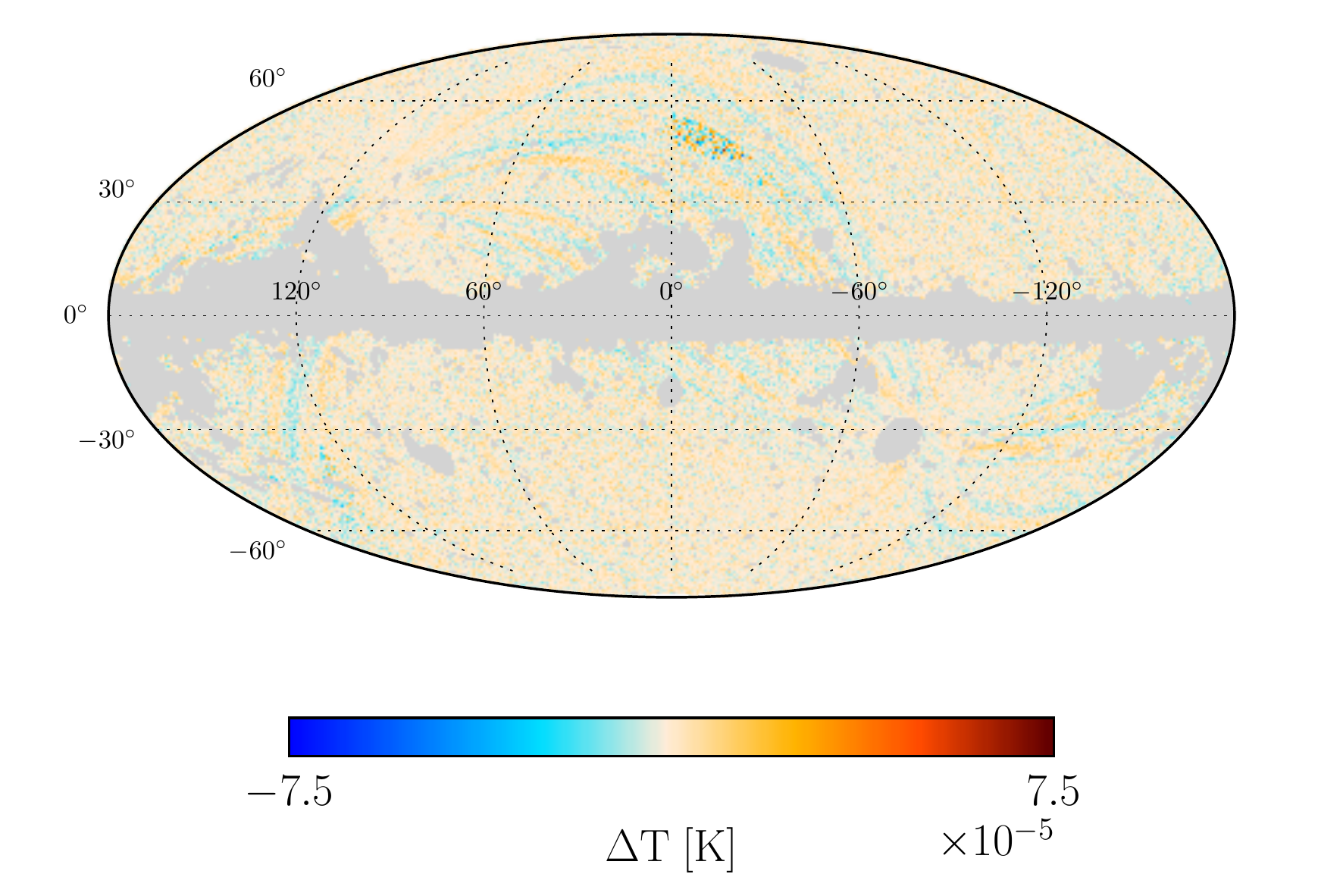}
\caption{Half-mission half-difference (HMHD) $\tt{Commander}$ CMB temperature anisotropy map. This map contains only noise and potential residual systematics.}
\label{fig:cmbnoisemap}
\end{figure}

\begin{figure}
\includegraphics[scale=0.6]{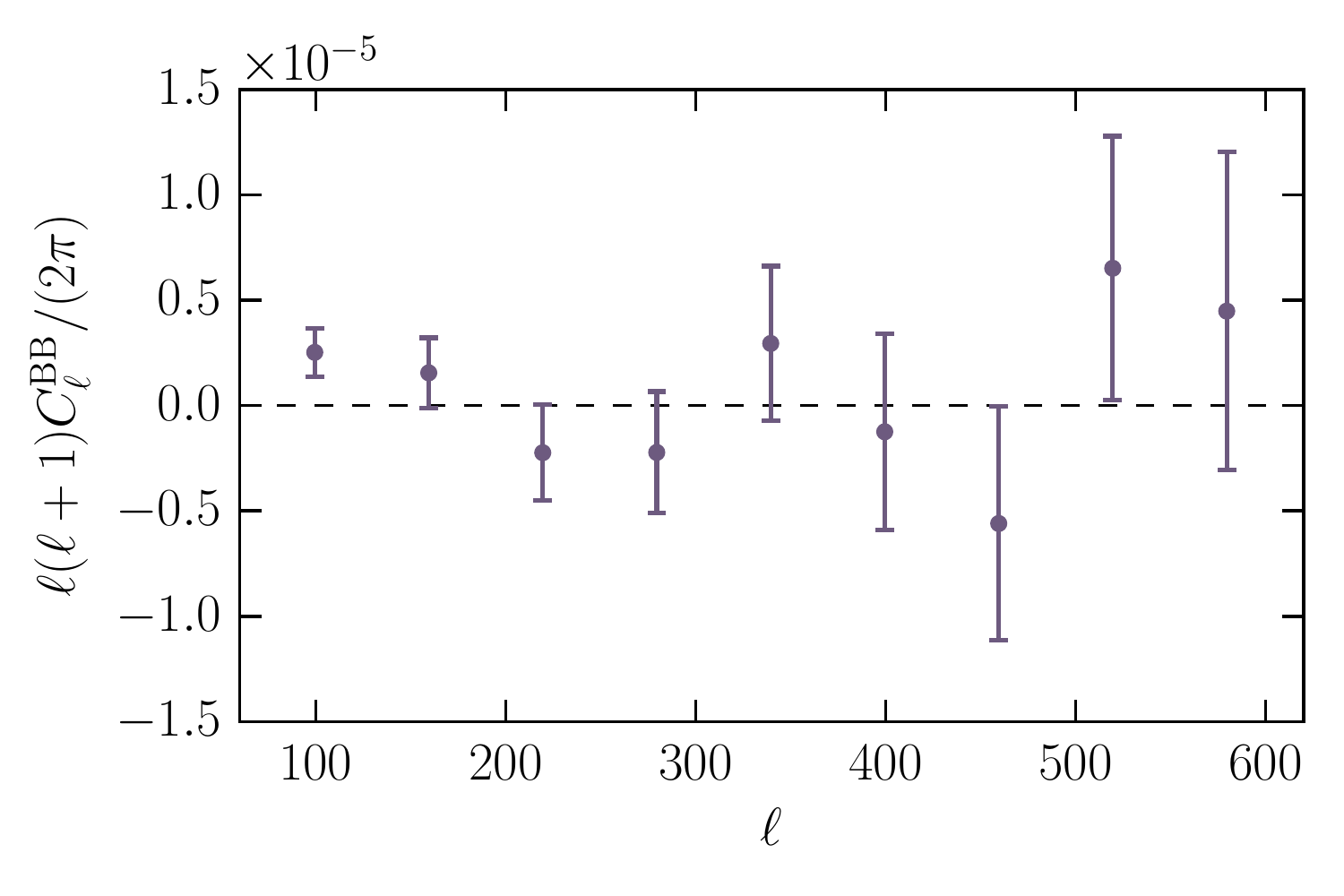}
\caption{Spherical harmonic power spectrum of cosmic shear B-modes computed from the SDSS Stripe 82 maps in Galactic coordinates. The angular multipole range and binning scheme is summarised in Table \ref{tab:clparams}.}
\label{fig:shearclsb}
\end{figure}

\bibliography{main_text_incl_figs}

\end{document}